\documentclass[10pt, preprint]{aastex}

\usepackage{graphicx}
\usepackage{amssymb}

\begin{document}

\title{Interactions of a Light Hypersonic Jet with a Non-Uniform Interstellar Medium} 

\author{Ralph  S. Sutherland and Geoffrey  V. Bicknell}
\affil{Research School of Astronomy and Astrophysics, \\ 
  Australian National University, ACT 0200, Australia.}

\shortauthors{Sutherland \& Bicknell}

\begin{abstract} 
We present three dimensional simulations of the interaction of a light hypersonic jet with an inhomogeneous thermal and turbulently supported disk in an elliptical galaxy. These simulations are applicable to the GPS/CSS phase of extragalactic radio sources. The interstellar medium in these simulations consists of a conventional hot ($T \sim 10^7 \> \rm K$) component together with a warm ($T \sim 10^4 \> K$) turbulently supported disk whose local density is described by a log-normal density distribution and whose spatial structure is realized from a fractal power-law.  We model the jet as a light, supersonic non-relativistic flow with parameters selected to be consistent with a relativistic jet with kinetic power just above the FR1/FR2 break.  

We identify four generic phases in the evolution of such a jet with the inhomogeneous interstellar medium: 1) an initial ``flood and channel'' phase, where progress is characterized by high pressure gas finding changing weak points in the ISM, flowing through channels that form and re-form over time, 2) a spherical, energy-driven bubble phase, were the bubble is larger than the disk scale, but the jet remains fully disrupted close to the nucleus, 3) a subsequent, rapid, jet break--out phase the jet breaks free of the last obstructing dense clouds, becomes collimated once more and pierces the spherical bubble, and 4) a classical phase, the jet propagates in a momentum-dominated fashion similar to jets in single component hot haloes, leading to the classical jet -- cocoon -- bow-shock structure. 

Mass transport in the simulations is investigated, and we propose a model for the morphology and component proper motions in the well-studied Compact Symmetric Object 4C31.04.  
\end{abstract}

\section{Introduction}
\label{s:intro}

There is a substantial literature on jet simulations going back to the time when the first supercomputers became available for computational astrophysics \citep{norman82a}. The simulations of extragalactic radio jets that have been carried out typically have
axisymmetric or cartesian slab-jet geometries; a smaller number of three dimensional simulations have also been conducted. One of the most important features has generally been the assumption of a hot, tenuous and smoothly distributed ambient atmosphere. These simulations have been extremely informative, leading to significant insights into the physics of extragalactic jets. For example the production of jet knots and filamentary structure has been associated with non-linear development of Kelvin-Helmholtz modes \citep{norman84a,hardee92a}; slab-jet simulations have shown how the extragalactic dentist drill \citep{scheuer82a} may work in practice \citep{hardee90a} and simulations of jets in atmospheres with density gradients \citep{hardee91a} have shown how the stability/instability of jets is influenced by the gradients in the ambient atmosphere. More recently, \cite{krause03a} has mapped out a region of the parameter space of jet density ratio, $\eta$ and Mach number, $M$ and has also simulated the progress of a light jet through a dense (but uniform) medium typical of Gigahertz Peak Spectrum (GPS) or Compact Steep Spectrum (CSS) radio sources. 

Notwithstanding the insights provided by the above work, until recently one key element has not been addressed in detail and this is the nature of the background medium. In the case of classical double radio sources, there is ample justification for assuming an homogeneous medium. However, there is excellent observational motivation for the disruption or modification of radio source morphologies by an \emph{inhomogeneous} medium. For example, the existence of filamentary optical line-emitting gas adjacent to the inner radio  lobes in M87 has been known for some time \citep{ford79a} as has the hierarchical structure of M87 evident in both radio and X-ray images \citep{owen00a,bohringer95a}. Moreover, recent Chandra observations of the inner few kpc of M87, indicate significant interaction of the inner radio plasma in M87 with the hot X-ray gas leading to the production of X-ray filaments.

There is also evidence for jet--ISM interaction in high redshift radio galaxies such as 4C41.17 \cite{bicknell00a}, quasars such as 3C~48 \citep{wilkinson91a}, blazars such as MKN~501 \citep{giroletti04a,bicknell05b}, and the general class of GPS and CSS sources \citep{devries99a,bicknell97a}, which motivated our current line of research in the first place. All of these sources show clear evidence for the interaction of a jet with a clumpy medium which substantially distorts the morphology of the radio source, in most cases producing accompanying emission line luminosity from the disturbed dense gas. 

There are other compelling reasons for considering the interaction of radio jets with an inhomogeneous medium. Ever since the paper of \citet{silk98a}, proposing an explanation for the \citet{magorrian98a} relation between black hole mass and bulge mass, there has been an increasing appreciation of the importance of feedback processes involving active galactic nuclei and the evolving interstellar medium of forming galaxies. In principle, jets, winds and radiation emitted from the environs of the black hole can impede continued accretion into a forming galaxy thereby limiting its growth. The importance of jets stems from the realisation that disrupted jets can process $4\pi$ steradians of solid angle in the interstellar medium (e.g. \citet{saxton05a}) and that they transport a large amount of momentum. For similar reasons the importance of jet -- ISM interactions has been appreciated in attempts to resolve the questions posed by cooling flows as well as the X-ray cavities associated with radio bubbles that are now being observed in these environments (e.g. \citet{nulsen05a}). 

In recent work we have begun to investigate these phenomena through the simulation of two-dimensional slab jets issuing into an inhomogeneous medium \citep{bicknell03a,bicknell03b,saxton05a}. We have shown that a numerous different radio morphologies can arise as a result of the interaction of both the jet and lobes with a clumpy interstellar medium and we have compared the resulting radio morphologies with a number of well--known radio sources. Some preliminary three-dimensional simulations have also been published \citep{sutherland05a,bicknell06a}.

In this paper we present a high resolution simulation of a radio jet interacting with an inhomogeneous interstellar medium in the form of a turbulently supported disk which extends our previous work in several different directions: (1) The simulation is three dimensional with a resolution of 512 cells in the (cartesian) coordinate directions and a spatial scale of 2 parsecs per cell. (2) The density structure of the warm disk is described by a log-normal distribution, typical of the warm interstellar medium in a number of different environments. (3) The initial distribution of the warm medium is that of a thick, almost Keplerian disk supported by a combination of thermal pressure and a, dominant, supersonic velocity dispersion. This type of distribution is one of several that may be contemplated and is supported by the observation of such disks in M87 \citep{dopita97a} and NGC~7052 \citep{vandermarel98a}. (Other types of distributions of warm gas that may be contemplated include that typical of the Lyman-$\alpha$ haloes observed in many high redshift radio galaxies. These will be considered in future papers in this series.) (4) The hard and soft X-ray emissivity and surface brightness of the thermal plasma is calculated, in conjunction with the radio surface brightness image. These synthetic images provide valuable insights for the interpretation of observational data.  

This simulation illustrates four important phases in the evolution of a young radio galaxy: (1) An initial ``flood and channel" phase wherein the radio source is beginning to be established, but is still interacting strongly with the disk material. (2) A quasi-spherical jet-driven bubble phase where the jet is fully disrupted and drives a pseudo-spherical bubble into the surrounding medium.  (3) a rapid jet establishment phase where the last obstruction is ablated away and the jet reforms and crosses the spherical bubble rabidly until it breaks out.  (4) A classical jet -- cocoon -- bow-shock phase familiar from previous studies of radio sources in a single component, hot ISM.  We also determine the X-ray morphology associated with the interaction of the radio source with the disk and show that the luminosity from the disk may persist. 

We begin, in the following section, by describing the key elements and assumption in some detail.  We  discuss our justification of, and constraints on, the parameters chosen for the simulations.

\section{The Jet-Galaxy System: Key Model Element and Parameters}

\subsection{Host galaxy potential}
\label{s:potential}

Elliptical galaxies contain both baryonic and dark matter. The former dominates on scales close to the core and the latter dominates at large radii, say of order 10~kpc.  Since jet simulations frequently extend over this range of scales, we have constructed a family of potentials which we use to represent a combined, self-consistent  distribution of baryonic (luminous) and dark matter. In the simulation presented here, the scale of the simulation is such that the potential is dominated by the baryonic component close to the core. However, the dark matter does have some influence. Hence, we present here a generic potential that is useful here and which can also be used in future simulations with a larger overall spatial scale, in which the dark matter has an even greater influence. The potential is based on isothermal distributions of baryonic and dark matter, described in terms of isotropic distribution functions, as follows.

 Let, $(\sigma_{\rm B}, \rho_{\rm B,0})$ and $(\sigma_{\rm D}, \rho_{\rm D,0})$ be the line of sight  velocity dispersions and central densities of baryonic and dark matter respectively and let the total specific energy of a baryonic particle (star, gas), or dark matter `particle',  be $E=1/2 v^2 + \phi_G$ where $v$ is the velocity and $\phi_G$ is the gravitational potential. The central value of the potential is taken as, $\phi_G(0) = 0$. We assume Maxwellian distributions for the distribution functions, for both dark and baryonic matter:
\begin{eqnarray}
f_{\rm D} (E) &=& \frac {\rho_{\rm B,0}}{(2 \pi)^{3/2} \sigma_{\rm B}^3} 
\> \exp (-E/\sigma_{\rm D}^2) \, ,\nonumber \\
f_{\rm B} (E) &=& \frac {\rho_{\rm D,0}}{(2 \pi)^{3/2} \sigma_{\rm D}^3} \>
\exp (-E / \sigma_{\rm B}^2) \, .
\end{eqnarray}
As for the classic single component isothermal sphere ({\em c.f.} \citet{king66}), we define dark and baryonic matter core radii, $r_{\rm D}$ and $r_{\rm B}$, via the relations:
\begin{equation}
\frac {4 \pi G \rho_{\rm D,0} r_{\rm D}^2}{\sigma_{\rm D}^2} =
\frac {4 \pi G \rho_{\rm B,0} r_{\rm B}^2}{\sigma_{\rm B}^2} = 9 \, .
\label{e:core_radii}
\end{equation}

For $\kappa  =  \sigma_{\rm D}/\sigma_{\rm B} \gg 1$ and $\lambda = r_{\rm D}/r_{\rm B} \gg 1$, the density is dominated by the baryonic component near $r=0$ and by the dark component near $r = r_{\rm D}$.   In this case, the core radii have the conventional interpretation of being the radii at which the surface densities of the respective components drop to approximately half of their central values.

Taking $\varrho =\rho_{\rm B,0}/\rho_{\rm D,0}$, the  dark and baryonic matter densities are given in terms of the dimensionless potential $\psi = \phi/\sigma_{\rm D}^2$ by:
\begin{eqnarray}
\frac {\rho_{\rm D}}{\rho_{\rm D,0}} &=& \exp (-\psi) \, , \nonumber \\
\frac {\rho_{\rm B}}{\rho_{\rm B,0}} &=& \frac {\rho_{\rm B,0}}{\rho_{\rm D,0}} \, 
\exp \, \left[ - \left( \frac {\sigma_{\rm D}}{\sigma_{\rm B}} \right)^2 \psi \right] \\
           &=& \varrho \, \exp \, \left( -\kappa^2   \psi \right) \, .
\label{e:densities}
\end{eqnarray}
 
Defining the normalized radius by $r^\prime = r / r_{\rm D}$, the dimensionless version of Poisson's equation is:
\begin{equation}
\frac {d^2 \psi}{dr^{\prime 2}} + \frac{2}{r^\prime} \frac {d \psi}{dr^\prime} =
9 \, \left[ \exp(-\psi) + \varrho  \exp \left( -\kappa^2 \psi \right)\right] \, .
\label{e:poisson}
\end{equation}

We may take the parameters of this double isothermal distribution to be $\kappa$, the ratio of the dark to baryonic velocity dispersions, and $\varrho$, the central baryonic to dark matter densities. However, it is generally more convenient to take $\kappa$, and the ratio of core radii, $\lambda$, as defining parameters, use $\varrho = \lambda^2 / \kappa^2$, and find the central densities from equations~(\ref{e:core_radii}),
\begin{equation}
\rho_{\rm D,0} = \frac {9 \sigma_{\rm D}^2}{4 \pi G r_{\rm D}^2} \, ,
\qquad 
\rho_{\rm B,0} = \frac {9 \sigma_{\rm B}^2}{4 \pi G r_{\rm B}^2} \, .
\end{equation}
We refer below to the double potential function for given $\kappa$ and $\lambda$ parameters as $\psi_{\kappa, \lambda}$.

Equation~(\ref{e:poisson}) is numerically integrated for chosen parameters, tabulated and interpolated with cubic splines when required by the hydrodynamic code.   In the present modeling we use $\kappa = 2$, and $\lambda = 10$,  referring to the potential as $\psi_{2,10}$.  The central density is dominated by baryonic material over dark matter by a ratio of $25:1$.   Figure \ref{f:pots} in the next section shows the resulting equilibrium  density distribution of the hot galactic atmosphere in the $\psi_{2,10}$ potential, compared to two single isothermal potentials.

\subsection{The Hot Galactic Atmosphere}

Generally, in radio galaxies, there are at least two components of the interstellar medium, a tenuous, smoothly distributed  hot atmosphere with central density $\sim 0.01 - 1 $ particles~$\rm cm^{-3}$ and $T \sim 10^7 \> \rm K$, and a cold/warm unevenly distributed  component with number densities $\sim 1-100 \> \rm cm^{-3}$ and temperatures $ \sim 10^{3-4} \> \rm K$. 
 
The tenuous, hot interstellar medium in our simulations is isothermal, although it is relatively straightforward to relax this condition by adopting, for example, an empirically determined radial temperature profile. Let $T$ be the gas temperature. The virial temperature corresponding to the dark matter is defined by $T_* = \bar{m}  \sigma_{\rm D}^2 / k$, where $ \bar{m}  = \mu m_u \approx 1.035\times 10^{-24}$~g, is the mean particle mass (including electrons, hydrogen and heavier ions),  using $\mu = 0.6224$, the mean molecular weight in fully ionised solar metallicity gas, and $m_u$, the atomic mass unit.  Assuming hydrostatic equilibrium, the particle number density, $n_{\rm h}$ of the hot gas is given simply by:
\begin{equation}
\frac {n_{\rm h}}{n_{\rm h,0}} = \exp \left[ - \frac {T_*}{T} \psi  \right] \, .
\end{equation}
(Note that this defines the total particle density, {\em not} the Hydrogen number density, often denoted by $n_{\rm H}$).  
On the 1~kpc scales simulated here,  the sound crossing time for the region at 400~kms$^{-1}$ is about 2.5~Myr, so the hydrostatic formalism used requires the atmosphere to have settled into place for longer than this prior to the jet interaction, which we assume to be the case.

\begin{figure}
\begin{center}
\includegraphics[width=5in]{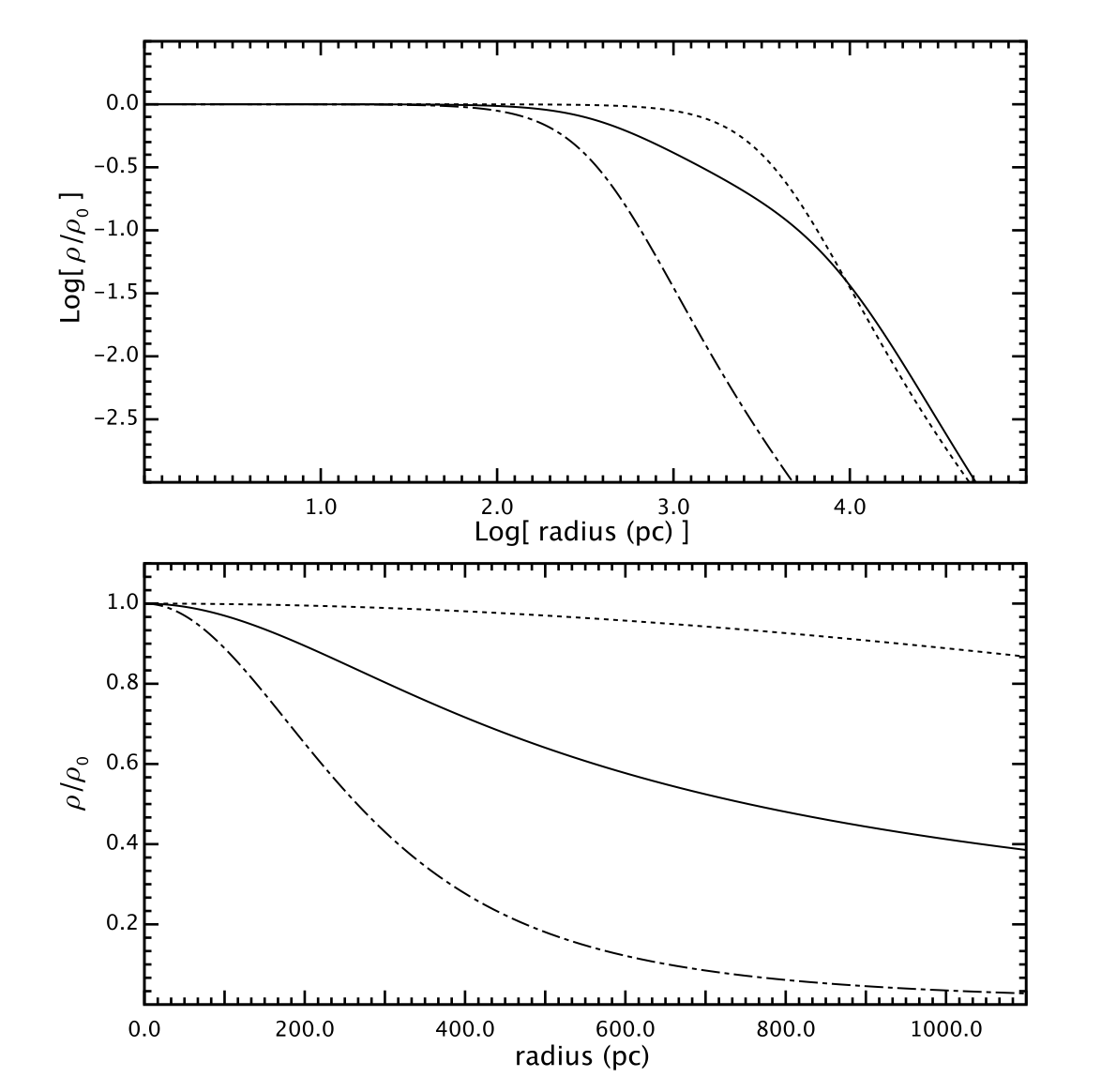}
\caption{{\footnotesize The density distribution of the normalized two component isothermal potential, $\psi_{2,10}$.  The solid curve represents the normalized total density distribution $\rho/\rho_0 = \exp[\psi_{2,10}]$, scaled to a dark matter radius of $r_{\rm D} = 3.5$~kpc, implying a baryonic core radius of $r_{\rm B} = 350$~pc.   The short dashed curve is the density distribution corresponding to an isothermal potential with a core radius of 3.5~kpc.  The dash-dot curve is a density distribution for an isothermal sphere with a core radius of 350~pc.  The upper panel uses logarithmic coordinates, the lower panel is linear and covers the domain of the simulated hot atmosphere.  See text for details. }}
\label{f:pots}
\end{center}
\end{figure}

Figure \ref{f:pots} shows how the $T = T*$ hot atmosphere fills the double isothermal, $\psi_{2, 10}$, potential with $r_{\rm D} = 3.5$~kpc.  For comparison, corresponding  functions  for single isothermal potentials, one each for a core radius of $r_{\rm c} = 3.5$~kpc, and $r_{\rm c} = 350$~pc, are overplotted.   The upper logarithmic plot shows the large scale behavior, and the lower linear panel focuses on the 1~kpc range used in the simulations here.  The key feature is that between the inner radius and outer radius, the density decreases with radius with an intermediate slope.  Within the 1~kpc range on the grid, the hot atmosphere density has fallen to approximately 40\% of its central value, and is essentially uniform (within 10\%) inside 200~pc, and is described by a radial powerlaw with a slope of index $-0.68$ in the outer regions to 1.0~kpc.

\subsection{ The non-uniform, fractal, warm interstellar medium}
\label{s:warm_ISM}

The warm and cold interstellar medium has been known to be non-uniform ever since the first observations of nebulae in the late 19th and early 20th centuries (although recognition of an interstellar medium as such came later in the 20th century).  In these simulations we introduce a  non-uniform medium, which is describable semi-analytically, with the view to assessing qualitatively the influence on inhomogeneity on the energetics of dynamical interactions -- even if a full theoretical understanding of the non-uniformity unavailable.  We compare a homogeneous model with an inhomogeneous model, at two resolutions, in order to comprehend  the consequences of non-uniformity will take, in a global dynamical sense.

To establish non-uniform medium we make use of an obvious analogy with a turbulent medium.  The physics literature on turbulence is vast, with results and models in a many fields of physical sciences, and astrophysics is is no exception.  Rather than attempt to review this huge topic here however, we refer the reader to the recent astrophysically oriented annual reviews of \citep{elmegreen04}, and \citep{scalo04} as a starting point for background material, and then refer simply to some specific results that we use below to arrive at reasonable (although not necessarily unique) parameters for our non-uniform medium.  

It must be emphasized that for the present work we are not modelling actual turbulence in a genuine causally generated ISM (see \citet{kritsuk06} for a recent large scale isothermal ISM turbulence simulation).  Rather, we are taking a parameterisation for the non-uniform properties of generic turbulent media focusing on characteristics such as the variance, $\sigma^2$, the intermittency (in skewed distributions), and {\em two--point} self-similar power-law structures, relying on a range of previous experimental and theoretical results from the field of turbulence. Hence the initial distribution of the ISM that we employ should be regarded as a physically motivated generalisation of a homogeneous model, whilst not necessarily representing an accurate  physical model of a turbulent ISM.

\subsubsection{Log-normal density distribution}
\label{s:lognormal}

Turbulence naturally gives rise to non-uniform structure, in velocity and density fields.  We neglect the velocity structure, and focus on choosing statistical parameters to describe the density.  This is justified numerically by the relatively small turbulent velocities expected when compared to the very large velocities found in our global jet--ISM interaction.  The velocities observed in typical warm ISM conditions fall in a range of transonic to mildly supersonic values, Mach 1-4 ( {\em e.g.} \citet{heiles04}).  At temperatures at or below $10^4$~K this corresponds to velocities $< 50 \> \rm km \> s^{-1}$.  In our 1~kpc simulations over $10^5$~yr timescales, the resulting displacements amount to less than 2 or 3 cells, and are insignificant compared to those of the radio jet and the energy bubble/cocoon generated by the main outburst, wherein velocities of hundreds or thousands of km/s occur.  Consequently, we do not impose a turbulent velocity field on the warm medium, and focus instead on constructing the density field.

We use a log--normal distribution to describe the single point statistics of the density field of our nonuniform ISM. The log--normal distribution is a skewed continuous probability distribution.  Unlike the normal distribution, it has a non-zero skewness, variable kurtosis, and in general the mode, median and mean are unequal.   The log-normal distribution appears to be a nearly universal property of isothermal turbulent media in experimental, numerical and analytical studies ( {\em e.g.} \citet{nordlund99a} , see also \citet{warhaft00}, and \citet{pumir94} ).   Moreover, it is encouraging  that the log--normal distribution is the limiting distribution for the product of random increments, in the same way that the normal distribution plays that role for additive random increments. It is thus compatible, at least conceptually, with a generic cascading process consisting of repeated folding and stretching.

We begin by describing parameters for an inhomogeneous interstellar medium density field, which is on average isotropic. That is, there is no dependence on location.   In \S~\ref{s:turb_disk} below, we describe how this standard distribution is modified to reflect the potential.  With a log--normal distribution, the natural logarithm of the ISM density field is a Gaussian which has a mean $m$ and variance $s^2$. The probability density function for the log-normal distribution of the mass density $\rho$  is,
\begin{equation}
P(\rho) = \frac{1}{s \sqrt{2 \pi} \, \rho} \exp \left \lbrack \frac{-(\ln \rho - m)^2}{2s^2}\right \rbrack \, .
\end{equation}
The mean $\mu$ and variance $\sigma^2$  of the density are given by
\begin{eqnarray}
\mu & = & \exp[ m + s^2/2] \,. \\
\sigma^2 & = & \mu^2 \, (\exp[ s^2]-1) 
\label{e:log_normal_pars}
\end{eqnarray}
Additional statistical properties of the log-normal distribution are summarized in Appendix~\ref{a:lognormal}.

In this simulation we adopt $\mu = 1.0$, $\sigma^2 = 5.0$, as our standard log--normal distribution. These values are compatible with the favored ranges in \citet{fischera03} and \citet{fischera04} from star burst galaxy reddening and extinction considerations.  The  variance measures how concentrated the mass is in dense cores, or conversely how much volume is occupied by voids, and these parameters give a flatness parameter $F = 1836$, indicative of an intermittent distribution (see Appendix~\ref{a:lognormal}).  With these parameters, densities below the mean, $z = \mu$, comprise one quarter of the mass, and  occupy three quarters of the volume, and the mean is approximately 20 times the mode (see \S~\ref{a:turb_disk} for further details).  Other values are possible of course, and a proposed  relationship between the log-normal variance and the isothermal Mach number of the turbulence given by \citep{nordlund99a}, viz.  
\begin{equation}
\sigma^2 \approx 0.25 M^2 \, ,
\label{e:varmach}
\end{equation}
suggests that higher mach numbers in an AGN medium for example, could be compatible with larger variances.  Below, when we use turbulent velocity support to determine the scale height of the warm ISM disk in the galactic potential, we adopt a turbulent velocity that is consistent with the value $\sigma^2 = 5.0$ adopted here; hence the turbulent velocity is not a completely independent parameter in the model.

\subsubsection{Power-law density structure}

The two--point structure of a homogeneous turbulent medium is best described in Fourier space.   We denote the Fourier transform of the density $\rho(r)$ by $F(k)$ (where $k$ is the wavenumber vector). The isotropic power spectrum $D(k)$ is the integral over solid angle in Fourier space of the spectral density $F(k) F^*(k)$. In three dimensions:
\begin{equation}
D(k) = \int  k^2 F(k) F^*(k) \> d \Omega .
\label{e:Ek}
\end{equation}. 
Even if the spectral density is anisotropic, the angular integral averages the spectral density into a one dimensional function of $k$ only.  

For a power-law dependence on $k$, $D(k) \propto k^{-\beta}$ and $\beta = 5/3$, the spectrum is referred to as Kolmogorov turbulence.  It has been shown that a scalar tracer (density) of the turbulent field also shows the Kolmogorov structure index \citep{warhaft00}. 

We follow \citet{fischera03} and adopt a standard density power spectrum with a Kolmogorov power-law with $\beta = 5/3$ to generate a spatial structure power-law for the density in our non-uniform ISM.  Given our assumption above that the (isothermal) disk is mainly supported by supersonic turbulence, a slightly different value of $\beta = 2.0$ associated with shock turbulence may be a reasonable alternative (e.g. \citet{boldyrev04a}) but we postpone investigation of the variation of $\beta$ to future work.
  
\subsubsection{Iterative Generation and Fractal Resolution}
 
For our inhomogeneous ISM we select $\mu = 1.0$ and $\sigma^2 = 5.0$ as our standard log--normal density field parameters, and a Kolmogorov $\beta = 5/3$ power-law spatial index.  This structure is modified as described below in \S\ref{s:turb_disk} to take account of a spatially variable mean density. The selected parameters may be representative in view of the results of \citet{fischera03} and \citet{fischera04} noted above. However, at this stage not a lot is known about the parameters of the inhomogeneous ISM of radio galaxies.  

In order to simultaneously achieve log-normal single-point statistics and a power-law self-similar structure, we have implemented the practical method developed by atmospheric scientists, for constructing two and three dimensional 
terrestrial cloud models, which are used in radiative transfer calculations \citep{lewis02}. The first step of the procedure involves constructing a cube in which each cell has a Gaussian distribution of mean zero. This cube is Fourier transformed and then apodized by a power-law in wave number. The apodized cube is then Fourier transformed back to the spatial domain and retains its Gaussian statistics because a sum of Gaussians is still a Gaussian. A cube with log-normal statisitics is then produced by exponentiating the Gaussian cube. However, as re-transformation to Fourier space shows this alters the power-law in wave number property.  Essentially the generation of dense cores and large voids consistent with the log-normal parameters, alters the low and high wavenumber ends of an original power-law, breaking self-similarity. Hence, the deviation from a power-law is calculated and this is used to estimate a correction to the power-law of the Gaussian distribution. Successive corrections are applied until satisfactory convergence to a power-law (within approximately 1\%) is obtained, usually in about 4--6 iterations for distributions with modest variances ($\sigma^2 < 10$). The process converges more slowly as the target variance, $\sigma^2$, increases, and variances $> 10$ were not attempted. The reader is referred to \citet{lewis02} for further details of this method.

From this process a library of cubes with a range of resolutions
and variances were pre-computed for use in a range of simulations, 
from which our standard set : $\mu = 1.0$ and $\sigma^2 = 5.0$, $\beta = 5/3$ was selected.

The remaining choice in this procedure is to select the range of wave numbers over which to generate the fractal, in particular the minimum wave number $k_{\rm min}$, which determines the largest structure scale in the resulting fractal with respect to the spatial grid.  With the scale height of the disk in the simulations being of order 100~pc in a domain of 1~kpc extent, a minimum wave number of $k_{\rm min } = 20$ is used to ensure that the largest `clouds' are of order 50~pc, and appropriate to the scale of the disk everywhere.  In the final structure used to form the disk, the wave numbers in the Fourier domain range from $k = 20$ to the Nyquist limit, $N/2-1 = 255$, covering just over 1.1 decades of scale.  This is limited by the computational requirements for the overall grid, and the fact that the disk is only a small part of the domain.  Significant improvement requires a substantial increase in computing resources than are available to treat more than say 2 decades of structure in the fractal disk alone.

That said however, we did perform the simulation with two resolutions, low and high, discussed in \S\ref{s:simpars}, to investigate resolution dependent differences, and the differences of the low and high resolution fractal simulations from a uniform model. The evolution of these different simulations is evidence that at least some of the fractal properties are being captured.   

\subsubsection{Equilibrium turbulent disks}
\label{s:turb_disk}

As noted in \S\ref{s:intro} many radio galaxies exhibit a turbulent disk of gas in the central regions, motivating us to consider the interaction of a jet with such a disk. Let us now consider the establishment of a  disk-like distribution of gas in some detail.

There are two elements to establishing a turbulent disk. We begin with the fractal, power-law distribution described above. This may be scaled to give a fractal distribution with a specific, but constant, mean density. This is unsatisfactory since the distribution of gas would not reflect the potential of the galaxy. Hence, we derive below expressions for the \emph{mean} density of a turbulent gas disk in the potential discussed in \S\ref{s:potential}. We use the spatially dependent mean density to scale the fractal cube. Essentially this provides a single realization from an ensemble of turbulent fractal disks.

Let the density of the warm disk gas be $\rho$ and its velocity be $v_i$. We express the density and velocity as statistical averages, as follows, with the angle brackets expressing ensemble averages:
\begin{equation}
\begin{array}{r c l r c l}
\rho &=& \bar \rho + \rho^\prime & \langle \rho^\prime \rangle &=& 0 \, ,\\
v_i &=& \tilde v_i + v_i^\prime & \langle \rho v_i^\prime \rangle &=& 0 \, .
\end{array}
\end{equation}
For a recent description of this mass-averaged approach to a statistical description of turbulent flow, see \citet{kuncic04a}.  We derive the relevant equations for our warm clumpy disk in the appendix, showing en-route that the azimuthal velocity in the turbulent disk is a function of the cylindrical radius only.

Following, \citet{strickland00a} we also adopt the ansatz of an {\em almost} Keplerian disk. Let $v_K = (r \, \partial \phi(r,0)/\partial r)^{1/2}$ be the Keplerian velocity in the disk mid-plane and put
\begin{equation}
\tilde v_\phi = e_K \, v_K = e_K \, \sigma_D \, 
\left(\frac {r \, \partial \psi (r,0)}{\partial r} \right)^{1/2} \, ,
\end{equation}
where $e_K$ is a constant close to unity. 

Taking the mean temperature of the disk to be $\tilde T$ and the line-of-sight turbulent velocity dispersion to be $\sigma_t$, the mean density of gas in the potential is given by:
\begin{equation}
\frac {\bar \rho(r,z)}{\bar\rho(0,0)} = 
\exp  \left[ - \frac {\sigma_D^2}{\sigma_g^2}
\left[ \psi(r,z) - e^2 \psi(r,0)- (1-e^2) \psi(0,0)  \right] \right] \, .
\label{e:warm_gas}
\end{equation} 
where $\sigma_g^2 = \sigma_t^2 + k \tilde T / \mu m$ and as previously 
$\sigma_D$ is the velocity dispersion of the dark matter. (Equation~(\ref{e:warm_gas}) assumes that $\tilde T$ and $\sigma_t$ are constant throughout the disk.) 

The main differences from the development of a similar equation by \citet{strickland00a} is the (mandatory) lack of dependence of $e_K$ on $z$ and the formal introduction of a turbulent velocity. The latter avoids the difficulty of prescribing an unphysically large temperature in order to achieve a reasonable disk scale height.

In summary then, we use the double isothermal potential $\psi_{\kappa, \lambda}$with $\kappa = 2$, $\lambda = 10$, 
scaled by $r_D = 3.5$~kpc, and $\sigma_D = 400$~km/s, and a rotation parameter $e_{\rm K} = 0.93$, plus a velocity dispersion of the warm gas of $\sigma_g = 40$~km/s for the disk.   This velocity dispersion corresponds to an adiabatic Mach number, $M \sim 4$, which is approximately consistent with the log--normal variance parameter $\sigma^2 = 5.0$ we adopted for the density field, if the \citet{nordlund99a} relation (equation (\ref{e:varmach}) holds.

\section{Jet Parameters}
\label{s:jetpars}

We use a non-relativistic code for these simulations. Other aspects of the code are discussed in \S\ref{s:code}. The use of such a code to simulate phenomena that involve relativistic flow in parts of the grid, is not ideal. Nevertheless, a large part of the flow field is non-relativistic and is driven by the energy or momentum flux provided by the jet. Hence, we establish jet parameters in such a way that the energy flux of the non-relativistic jet corresponds to the jet energy flux of a given relativistic jet using relationships derived by \citet{komissarov96a}.
 
The relationships between the respective densities and Mach numbers are given by the following relationships for relativistic and non-relativistic jets with the same energy flux, velocity and pressure:
\begin{eqnarray}
\rho_{\rm nr} &=& \frac {2 \gamma}{\gamma-1} \, \frac {p}{c^2} \, 
\Gamma^2 \left[ 1 + \frac {\Gamma}{\Gamma+1} \chi \right] \, . \\
M_{\rm nr}^2 &=& \frac {2 M_{\rm rel}^2}{2-\gamma} \,
\left[1+  \frac {\Gamma}{\Gamma +1} \chi \right]
\left[ 1 + \frac {\chi}{2-\gamma} \right]^{-1} \, .
\end{eqnarray}
\citep{komissarov96a}.
We use these relationships to determine the  ratio of the non-relativistic jet density to the background and the non-relativistic jet Mach number. First note that the relativistic Mach number is given by:
\begin{equation}
M_{\rm rel}^2 = \frac{2-\gamma}{\gamma-1} \Gamma^2 \beta^2 \left[ 1 + \frac {\chi}{2 - \gamma} \right] \, .
\end{equation}
As a result of the above relationships, the following equations for the conventional non-relativistic jet parameters, the density ratio, $\eta$ and the Mach number, $M_{\rm nr}$ are derived. The parameters $\xi$ and $T_{\rm ism}$ appearing in equation~(\ref{e:eta}) are the ratio of jet to external pressures  and the (hot) interstellar medium temperature respectively.
\begin{eqnarray}
\eta &=& \frac {2 \gamma}{\gamma-1} \xi \left( \frac {kT_{\rm ISM}}{\mu m c^2} \right) \Gamma^2 \left[ 1 + \frac {\Gamma}{\Gamma+1} \chi \right] \, .\label{e:eta}\\ 
M_{\rm nr}^2 &=&\frac {2}{\gamma-1} \left[ 1 + \frac {\Gamma}{\Gamma+1} \chi \right]
\left( \Gamma^2 - 1 \right) \, .
\label{e:M_nr}
\end{eqnarray}
Note that the low value of $kT/\bar{m}c^2 \sim 10^{-6}$ guarantees a light non-relativistic jet ($\eta \ll 1$) despite a high Lorentz factor, and that  the non-relativistic Mach number effectively corresponds to Lorentz factor.

In our simulations we take the gas to have the ideal adiabatic index, $\gamma=5/3$, since this represents the external medium the most accurately and we are mainly interested in the effect that the jets have on the external clumpy medium.

The following expressions for the (equivalent) versions of the jet power are also useful. Let $D$ represent the diameter of the jet with cross-sectional area $A= \pi D^2/4$. Then, the relativistic and non-relativistic jet powers are given by:
\begin{eqnarray}
F_{\rm E, rel} &=& \frac {\gamma}{\gamma-1} \,
c p_{\rm jet} A \Gamma^2 \beta \,
\left[ 1 + \frac{\Gamma-1}{\Gamma} \chi \right] \, , 
\nonumber \\
&=& 3.9 \times 10^{40} \, \frac {\gamma}{\gamma-1} \,
\xi \, \left( \frac {p_{\rm ism}/k}{10^7} \right) \,
\left( \frac {D}{10 \rm pc}\right)^2 \,
\Gamma^2 \beta \,
\left[ 1 + \frac {\Gamma-1}{\Gamma} \chi \right] \quad
\rm ergs \> s^{-1} \, . \label{e:fe_rel}\\
F_{\rm E,nr} 
&=& \frac {\gamma}{\gamma-1} \, p_{\rm jet} v A
\, \left[ 1 + \frac {\gamma-1}{2} M_{\rm nr}^2 \right]  \, ,
\nonumber \\
&=& 3.9 \times 10^{40} \, \frac {\gamma-1}{\gamma} \,
\xi \, 
\left( \frac {p_{\rm ism}/k}{10^7} \right) \,
\left( \frac {D}{10 \rm pc}\right)^2  \, \beta \,
\left[ 1 + \frac {\gamma-1}{2} M_{\rm nr}^2 \right] 
\quad \rm ergs \> s^{-1} \, .
\end{eqnarray}
A point to note from these equations, which is relevant to the choice of jet parameters, is that for a given jet Lorentz factor and the ratio $\chi$ of rest-energy density to enthalpy, the jet power is proportional to the 
ratio $\xi$ of jet pressure to ISM pressure times the ISM pressure. Even for a relatively high ISM pressure $\sim 10^7$ and a high Lorentz factor, it is likely that for a jet to achieve FR2 type powers well in excess of $10^{43} \> \rm ergs \> s^{-1}$ the ratio of jet to ISM pressures ($\xi$) is much greater than unity.

\subsection{Resolution Constraints on Jet and ISM parameters}
\label{s:resolution}

\subsubsection{Jet parameters}

Here we discuss the selection of parameters for the jet, potential and hot and warm components of the atmosphere. Physical objectives dominate the criteria by which we select these parameters but the selection is also governed by the desired resolution and spatial dynamic range constrained by the necessity for a realistic number of cells in the grid.
In this subsection we discuss other, global, resolution constraints on the simulation; in the following subsection (\S~\ref{s:scaling}) we consider how the simulations may be scaled.

If a jet is 20~pc wide where it enters the grid and we require 10 cells across the jet in order to resolve it adequately with a lagrangian--remap  PPM algorithm, then the maximum spatial size of a $512^3$ grid is just over a kiloparsec. Hence, using a code with fixed sized cells, such as \emph{ppmlr} the range of scales is limited to those relevant to Gigahertz Peak Spectrum (GPS) sources. Nevertheless, some of the features present in the simulations would probably also be relevant to larger scale sources and we indicate these in the sections below.

In adiabatic simulations the precise choice of parameters is not highly constrained since arbitary spatial, velocity and density scales may be applied. However, since we have introduced cooling processes the scaling that is allowed is restricted to a one parameter set (see \ref{s:scaling}) so that we need to exercise some care in selecting parameters that provide both physically consistent and interesting simulations, which reveal the range of feasible interactions between a jet and an inhomogeneous interstellar medium. The process of choosing realistic parameters is both interesting and informative.

For the jet the relevant relativistic parameters are: Velocity in units of the speed of light $\beta$ (equivalently Lorentz factor $\Gamma$), pressure $p_{\rm jet}$ defined by its ratio $\xi$ to the external ISM pressure and the proper density parameter $\chi = \rho_{\rm jet} c^2/4p_{\rm jet}$. The non-relativistic counterparts are the velocity $v$, Mach number $M$, the pressure, and the ratio of jet to ISM densities, $\eta$.

The kinetic power of a jet [ see equation~(\ref{e:fe_rel})] provides one constraint on jet parameters. Another useful constraint comes from consideration of the hot-spot advance speed. Many FR2 radio lobes display broad structure that indicates that the head of the lobe is not expanding at high velocity; this is confirmed by statistical estimates of average lobe expansion in powerful sources $\sim $ a few percent of the speed of light \citep{scheuer95a}. (See also \citet{blundell99a}.) Following the idea of \citet{scheuer82a} which more recently has become formally expressed in the form of self-similar evolution of radio galaxy lobes (e.g. \citet{falle91a,begelman96a,bicknell97a}) the lobe advance is affected by the spread of the jet momentum over an area that is larger than the jet cross-section. On the other hand the ``instantaneous'' hot-spot velocities in a number of GPS sources have been observed to be a considerable fraction of the speed of light (e.g. \citet{conway02a}). Therefore in the first instance it is useful to consider the hot-spot advance speed $c \beta_{\rm hs}$ calculated from ram pressure balance at the jet terminus. For a relativistic jet expanding into a thermal medium, the instantaneous hot spot advance speed is given by:
\begin{equation}
\frac {\beta_{\rm hs}}{\beta} \approx
\frac {\alpha \Gamma}{1 + \alpha \Gamma} \, ,
\end{equation} 
where
\begin{eqnarray}
\alpha &=& \left[ \frac {\rho_{\rm jet} + 4 p_{\rm jet}/c^2}{\rho_{\rm ism}} \right]^{1/2} =
 \left( \frac {4 p_{\rm jet}}{\rho_{\rm ism} c^2} \right)^{1/2} \,
\left( 1 + \chi \right)^{1/2}  \, .
\end{eqnarray}
\citep{safouris06a}.

Let $n_{\rm ism}$ be the total number density of the interstellar medium in the vicinity of the hot spot and let $D$ be the jet diameter. Then, the jet kinetic power can be expressed in terms of the hot-spot parameters as follows:
\begin{eqnarray}
F_{\rm E} &=& \mu  m c^3  \, 
\left[ 
\frac {1+\frac{\Gamma}{\Gamma -1} \chi}
{1 + \chi} \right] \,
\left[ \frac {\beta_{\rm hs}^2/\beta^2}
{1 - \beta_{\rm hs}^2/\beta^2 } \right] \,
n_{\rm ism} \, \beta A  \, ,  \nonumber \\
&\approx& 2.1 \times 10^{46}
\left[ 
\frac {1+\frac{\Gamma}{\Gamma -1} \chi}
{1 + \chi} \right] \,
\left[ \frac {\beta_{\rm hs}^2/\beta^2}
{1 - \beta_{\rm hs}^2/\beta^2 } \right] \, n_{\rm ism} \,  \beta \,
\left( \frac {D}{\rm 10 \> pc} \right)^2 
\quad \rm ergs \> s^{-1}  \, .
\end{eqnarray}
An advantage of this expression is that the precise value of the parameter $\chi = \rho_{\rm jet} c^2 / 4 p_{\rm jet}$ is not important for moderate to high Lorentz factors. Most of the dependence of the jet power enters through the local interstellar medium density, the jet diameter and the instantaneous hot spot advance speed.

At a kiloparsec from the core, $D \sim 10 \> \rm pc$ and $n_{\rm ism} \sim 1$ are reasonable fiducial values. At 10~kpc, $D \sim 1 \> \rm kpc$ and $n_{\rm ism} \sim 10^{-2} - 10^{-3}$ may be more appropriate. In the first case, $\beta_{\rm hs} \approx 0.5$ and $\beta \approx 1$ give a jet power $\approx 7 \times 10^{45} \> \rm ergs \> s^{-1}$ typical of the most powerful FR2 jets; $\beta_{\rm hs} \approx 0.03$ implies a power $\sim 2 \times 10^{43} \> \rm ergs \> s^{-1}$ at the lower end of the FR2 range. On the other hand, for FR2 jets on a scale of 10 kpc, $\beta_{\rm hs} \approx 0.03$, $D \approx 1 \> \rm kpc$ and $n_{\rm ism} \approx 0.01 \rm cm^{-3}$ gives a power $\sim 2 \times 10^{45} \> \rm ergs \> s^{-1}$. Thus FR2 jets remain powerful from small to large scales if the effective diameter of the jet widens, in response to the evolving dynamics of the lobe (which confines the jet) and the instabilities in the jet itself leading to jittering and filamentation.

We have selected a jet for which the instantaneous 
$\beta_{\rm hs}$ is of order 0.05 - 0.1 and the initial jet diameter is $ 20 \> \rm pc$. These parameters, together with a central ISM  $p/k = 10^6 \> \rm cm^{-3} \> K$ places the jet at the low end of FR2 power $\approx 3 \times 10^{43} \> \rm ergs \> s^{-1}$. This choice of parameters produces a jet which initially interacts strongly with the ambient ISM but whose morphology at later times is similar to a classical FR2 radio source. Test simulations (not presented here) show that with higher-powered jets the hot spot advance rapidly becomes focused when the jet emerges from the inhomogeneous region surrounding the core and the advance speed is quite high. Thus, this simulation is designed to show the characteristics of the  range of interactions that can occur. However, for these characteristics to be be manifest in higher powered sources, it is probably necessary to relax the assumption of an equilibrium disk for the inhomogeneous ISM.  Future parameter studies will be used to map out different regimes in more detail.

Radio galaxies are mainly associated with the high luminosity end of the elliptical galaxy distribution. However, FR2 galaxies are less optically luminous than FR1s \citep{ledlow94a}. Therefore, we adopt a velocity dispersion of the baryonic matter of $200 \> \rm km \> s^{-1}$. The velocity dispersion of the dark matter is $400 \> \rm km \> s^{-1}$. The latter choice is driven by the physical requirement that the dark matter in an elliptical galaxy is more extended than the baryonic matter and by the numerical requirements imposed by resolution and number of cells discussed above.  This combination of different dark matter and baryonic matter scales is naturally treated by the double isothermal potential described in \S\ref{s:potential}.

\subsubsection{Hot ISM parameters}

The selection of parameters for the  hot medium is straightforward. We choose a temperature for the (isothermal) medium close to the virial temperature defined by the dark matter.  As noted above we specify the density by a value of  $p/k = 10^6 \> \rm cm^{-3} \> K$. The hot ISM distribution is defined first with the calculated density and pressure applied to each cell in the grid. An algorithm for the warm medium is then applied as described in the following paragraph; this algorithm replaces some of the cells in the hot ISM by warm gas.

The warm inhomogeneous interstellar medium is prescribed in the form of a turbulently supported disk. The aim is to realize a single instance out of the ensemble of possible distributions described by the log-normal, power-law energy spectrum distributions with a spatially dependent mean density described in \S\ref{s:warm_ISM}.
\begin{enumerate}
\item The disk is taken to be isothermal, with a temperature of $T_{\rm w} = 10^4 \> \rm K$, characteristic of the equilibrium temperature in an ambient radiation field. In order that a disk have a scale height that guarantees significant interaction with an emerging jet, the assigned turbulent line of sight velocity dispersion $\sigma_{\rm t}$ is supersonic. Observationally, this is justified by the inference of supersonic turbulence in the disk in the center of M87 \citep{dopita97a} and the gaseous disk in NGC~7052 \citep{vandermarel98a}. The density scale of the log-normal distribution at the centre of the galaxy, $\rho_{\rm w} (0,0,0)$ is defined by approximate total (thermal plus turbulent) pressure equilibrium with the ambient hot medium, that is,
\begin{equation}
\bar \rho_{\rm w} (0,0,0) \left( \frac {kT}{\bar{m}} + \sigma_t^2 \right) \approx 
p_{\rm hot} (0,0,0)  \, .
\end{equation}
\item The density scale of the warm gas at each cell of the simulation is defined by equation~(\ref{e:warm_gas}) for the mean density $\bar \rho_{\rm w}(r,z)$ of warm turbulent gas in the potential well of the galaxy. Let $f_\rho(i,j,k)$ represent the log-normal unit mean cube computed as described in \S\ref{s:lognormal}, with the triplet $(i,j,k)$ representing the cell indices. The density of warm gas assigned to each cell is:
\begin{equation}
\rho_{\rm w} (i,j,k) = \bar \rho_{\rm w} (x_c,y_c,z_c) \times f(i,j,k)  \, ,
\label{e:warm_density}
\end{equation}
where $(x_c,y_c,z_c)$ are the zone-centred coordinates corresponding to 
$(i,j,k)$.
\item There is a cutoff applied to each cell: Where the statistically distributed density falls below the ambient hot ISM density the hot ISM in that cell is not replaced. 
\end{enumerate}

\subsection{Scaling}
\label{s:scaling}

With adiabatic simulations, the choice of the spatial, velocity and density scales is arbitrary. However, the introduction of cooling (in the thermal gas) restricts the allowable scaling to a one-parameter set. 

We define scaling parameters and scaled variables, denoted by primes, through the following relationships. All variables have their usual meanings with $\rho$ being the gas density, $\rho_*$ the stellar plus dark matter density , $\rho^2 \Lambda_\rho (T) $ is the volume emissivity due to cooling (with $\Lambda_\rho (T)$ the density-based cooling function) and $\phi_{\rm G}$ the gravitational potential. 

\begin{equation}
\begin{array}{r c l r c l r c l}
x_i  &=& x_0 \, x_i^\prime  \, , & t &=& t_0 \, t^\prime  \, , \\
\rho &=& \rho_0 \,\rho^\prime  \, ,& p &=& p_0 \, p^\prime  \, ,& T &=& T_0 \, T^\prime  \, ,\\
\phi_{\rm G} &=& \phi_{\rm G,0} \, \phi_G^\prime  \, ,& \Lambda_\rho (T) &=& 
\Lambda_0 \Lambda_\rho^\prime (T_0 t^\prime)  \, .\\
\end{array}
\end{equation}

Scaling the continuity and momentum equations in such a way that the form of the equations is preserved, is straightforward and identical to the adiabatic case,  resulting in the scaling relationships and scaled equations:
\begin{enumerate}
\item General scaling,
\begin{equation}
v_0 = \frac {x_0}{t_0} \qquad \epsilon_0 = p_0 = \rho_0 \, v_0^2 \qquad 
\phi_{\rm G,0} = v_0^2   \, .
\end{equation}
\item The ideal equation of state $p=\rho k T /\bar{m}$ implies that
\begin{equation}
\frac {kT_0}{\bar{m}} = v_0^2  \, .
\end{equation}
\item The scaled internal energy equation:
\begin{equation}
\frac {d \epsilon^\prime}{dt^\prime} - h^\prime \frac {d \rho^\prime}{dt^\prime}
= - \frac{\Lambda_0 \rho_0^2 t_0}{\epsilon_0} \, {\rho^\prime}^2 \,
\Lambda_\rho^\prime (T_0 T^\prime)   \, .
\end{equation}
\item The scaling parameter for the cooing function:
\begin{equation}
\Lambda_0 = \frac {p_0}{\rho_0^2 t_0} = \frac{v_0^3}{\rho_0 x_0}  \, .
\label{e:lambda_0}
\end{equation}
\item Finally, the scaled form of the potential equation for the gravitational potential is
\begin{equation}
{\nabla^\prime}^2 \phi_{\rm G}^\prime = 4 \pi G^\prime \, \rho_*^\prime  \, ,
\end{equation}
where $G^\prime$ is defined in terms of Newton's constant of gravitation, $G$, and the gravitational mass density scale, $\rho_{*,0}$ by
\begin{equation}
G^\prime = \frac {G \rho_{*,0} x_0^2}{v_0^2} = \frac {G M_{0,*}}{x_0 v_0^2}  \, ,
\end{equation}
where $M_{*,0} = \rho_{*,0} x_0^3$. 
\end{enumerate}

In order that a given simulation describe a set of scaleable physical situations it is necessary that the scaled cooling function have the same functional form. Thus unless $\Lambda_\rho^\prime (T_0 T^\prime)$ has some special from (e.g. a power-law) it is necessary that the parameter $T_0$ be invariant under scaling. Hence, the parameter $v_0$ is invariant; this is the major difference from adiabatic scaling in which $v_0$ may be arbitrary. The spatial scale $x_0$ is arbitrary, but $t_0 = x_0/v_0$ is restricted by the constancy of $v_0$.

Moreover, in order that the primed equations describe the same situation, the parameter $\Lambda_0$ must be invariant. Referring to equation~(\ref{e:lambda_0}) this means that the gas density scale is inversely proportional to the spatial scale: $\rho_0 \propto x_0^{-1}$.

Invariance of the same scaled gravitational equations requires $G^\prime$ to be invariant so that $\rho_{*,0} \propto x_0^{-2}$. This means that the ratio of gas to gravitating density scales $\rho_0/\rho_{*,0} \propto x_0$. Hence, if we increase the physical scale of a simulation the ratio of physical gas to gravitating mass density increases. Clearly we cannot do this indefinitely since this would invalidate the neglect of self-gravity of the gas.

To summarize a given simulation defines a one-parameter family of simulations where the scaling parameters satisfy the following constraints:
\begin{equation}
\begin{array}{r c l r c l r c l}
x_0 &=& \hbox{arbitrary} ,& v_0 &=& \hbox{fixed} \, ,& t_0 &=&{x_0}/{v_0}\, , \\
\rho_0 &\propto & x_0^{-1} \, ,& p_0 &=& \rho_0 v_0^2 \, ,&{kT_0}/{\bar{m}} &=& v_0^2\, , \\
\rho_{*,0} &\propto & x_0^{-2}\, , & M_{*,0} &\propto & x_0\, .\\
\end{array}
\end{equation}
with the proviso that $x_0$ cannot increase to the extent that the gas density exceeds the density of gravitating matter.

Notwithstanding this restricted one-parameter scaling, the allowable set of physical models allowed by this scaling describes an interesting variety of different physical situations.

\subsection{Summary of simulation parameters}
\label{s:simpars}

\subsubsection{Model  {\bf A}}

Three simulations were performed  in this paper, models {\bf A}, {\bf B}, and {\bf C}. However, the first (Model {\bf A}) is presented as the standard model, in the greatest detail. Models {\bf B} and {\bf C} are comparison simulations designed to highlight the effect of the density and the distribution (smooth or fractal) of the warm gas.

Tables~\ref{t:Model_A_pars_jet}, \ref{t:Model_A_pars_pot} and 
\ref{t:Model_A_pars_Gas} summarize the parameters used in Model~{\bf A}, for the jet, the potential and the interstellar medium respectively. The original scaling uses a spatial scale of $x_0 = 1 \> \rm kpc$, which is the size of the grid; the parameters relevant to this scaling are given in the third column of the table. However, as shown above there is a one-parameter degree of freedom in the scaling that may be used and indicative sets of parameters are given for $x_0 = 0.2$ and 5~kpc in the fourth and fifth columns respectively.

\subsubsection{Models  {\bf B} and  {\bf C}}

In addition to the main simulation model~{\bf A}, models {\bf B} and model~{\bf C} were performed to look for changes in the interaction sequence with a change in two disk ISM parameters, density and uniformity.
 
The only physical parameter that is different in Model B is the mean density of warm gas. The central value in model {\bf B} is $20 \> \rm cm^{-3}$ compared to $10 \> \rm cm^{-3}$ in model~{\bf A}.  The third simulation, model {\bf C}, has the same mean density as model {\bf B}, but has a perfectly smooth, non-fractal distribution of warm gas. The purpose of model {\bf B} is to examine the effect of the mean density of identically distributed gas. The purpose of model {\bf C} is to examine the effect of the porosity of the gas distribution. In models {\bf A} and {\bf B} the distribution of warm gas is such that the jet plasma can force its way through low density channels. In model {\bf C} the only route for the jet to force its way into the hot interstellar medium is for it to push the dense gas out of the way.  

The resolution of Models {\bf B} and {\bf C} is a factor of two lower, {\em i.e.} $256 \times 256 \times 256$. However, the lower resolution does not appear to affect the comparison. We do not expect the much more uniform density distribution in model {\bf C} to suffer from lower resolution; in model {\bf B} we observe similar behavior to that of model~{\bf A} and the jet is adequately resolved with 10 resolution elements across its diameter.

\begin{table}
\begin{center}
{\small 
\begin{tabular}{l c c c c }
\hline
          &                & \multicolumn{3}{c}{Scaled to  $x_0 =$} \\
Parameter \& units       &    Symbol           & $1.0 \rm kpc$  &  $0.2 \rm kpc$  & $5.0 \rm kpc$ \\
\hline
\hline
\multicolumn{3}{l}{\bf Equivalent Relativistic  Jet Parameters} \\
{\dag Lorentz factor}              & $\Gamma$  &{5} &{5} &{5}      \\
{\dag Rest energy density/enthalpy}& $\chi$    &{10}  &{10}  &{10}     \\
{Velocity / Speed of light}        & $\beta$   &{0.9798}&{0.9798}&{0.9798} \\
\multicolumn{3}{l}{\bf Hydrodynamic  Jet Parameters} \\
{ Pressure / External pressure }  & $\xi$     & {1.0} & {1.0} & {1.0} \\
{ Density  / External density  }    & $\eta$    & {$2.0\times 10^{-3}$}  & {$2.0\times 10^{-3}$}  & {$2.0\times 10^{-3}$}   \\
{ Mach number  }                      & $M$       & {25.9} & {25.9} & {25.9} \\
{\dag Diameter} (pc)  & $D_{\rm jet}$      &40 & 8.0 & 200.0 \\
Kinetic luminosity    & $L_{\rm jet}$      &{$2.77\times 10^{43} $}& {$5.54\times 10^{42}$}& {$1.385\times 10^{44}$}\\
\hline
\multicolumn{5}{l}{{\footnotesize Assigned parameters are indicated with a \dag symbol; others are derived.}} \\
\end{tabular}
\label{t:Model_A_pars_jet}
}
\end{center}
\caption{Standard Jet Parameters of Model {\bf A}.}
\end{table}

\begin{table}
\begin{center}
{\small 
\begin{tabular}{l c c c c }
\hline
          &                & \multicolumn{3}{c}{Scaled to  $x_0 =$} \\
Parameter and units       &    Symbol           & $1.0 \rm kpc$  &  $0.2 \rm kpc$  & $5.0 \rm kpc$ \\
\hline
\hline
\multicolumn{3}{l}{\bf Double isothermal potential} \\
{\dag Dark matter to } \\
{ Baryonic velocity dispersion}& $\kappa $  & {2}& {2}& {2} \\
{\dag Dark matter to } \\
{ Baryonic core radius}        & $\lambda $  & {10} & {10} & {10} \\
{ Central Baryonic density to } \\
{ Dark matter density}   & $\varrho $  & {25}& {25}& {25} \\
\\
\multicolumn{3}{l}{\bf Dark matter values }\\
{\dag Velocity dispersion} 
($\rm km \> s^{-1}$) & $\sigma_{\rm d}$       &400 & 400 & 400\\
{\dag Core radius} 
(kpc)   & $r_{\rm d}$                         &3.5 & 0.7 &  17.5\\
{ Central density}  
($\rm g \> cm^{-3}$)    & $\rho_{\rm d,c}$    &$1.47\times 10^{-22}$ & $3.68\times 10^{-21}$ & $5.88\times 10^{-24}$ \\
\multicolumn{2}{l}{\bf Enclosed Masses, at $r = r_{\rm d}$ }\\
{ Dark Mass  }     ($M_\odot$)  & $M_{\rm d}(r_{\rm d})$  & $8.14\times 10^{10}$  & $1.63\times 10^{10}$  &  $4.07\times 10^{11}$ \\
{ Baryonic Mass }  ($M_\odot$) & $M_{\rm b}(r_{\rm d})$  & $4.56\times 10^{10}$  & $9.11\times 10^{9}$   &  $2.28\times 10^{11}$ \\
{ Total Mass  }    ($M_\odot$)  & $M_{\rm T}(r_{\rm d})$  & $1.27\times 10^{11}$  & $2.54\times 10^{10}$  &  $6.35\times 10^{11}$ \\
{ Baryonic/Dark Mass ratio } &$M_{\rm b}(r_{\rm d}) / M_{\rm d}(r_{\rm d})$ &0.6 & 0.6 & 0.6\\
\\
\multicolumn{3}{l}{\bf Baryonic values }\\
{ Velocity dispersion} 
($\rm km \> s^{-1}$) & $\sigma_{\rm b}$        & 200 & 200 & 200 \\
{ Core radius} 
(kpc)  & $r_{\rm b}$                           &0.35 & 0.07 & 1.75 \\
{ Central density} 
($\rm g \> cm^{-3}$)   & $\rho_{\rm b,c}$      &$3.68\times 10^{-21}$ & $9.20\times 10^{-20}$ &  $1.47\times 10^{-22}$ \\
\multicolumn{2}{l}{\bf Enclosed Masses, at $r = r_{\rm b}$ }\\
{ Dark Mass }    ($M_\odot$)   & $M_{\rm d}(r_{\rm b})$  & $3.08\times 10^{8}$   & $6.15\times 10^{7}$   &  $1.54\times 10^{9}$ \\
{ Baryonic Mass }   ($M_\odot$)& $M_{\rm b}(r_{\rm b})$  & $4.72\times 10^{9}$   & $9.44\times 10^{8}$   &  $2.36\times 10^{10}$ \\
{ Total Mass }  ($M_\odot$)  & $M_{\rm T}(r_{\rm b})$  & $5.03\times 10^{9}$   & $1.01\times 10^{9}$    & $2.51\times 10^{10}$ \\
{  Baryonic/Dark Mass ratio } &$M_{\rm b}(r_{\rm b}) / M_{\rm d}(r_{\rm b})$ &15.3 & 15.3 & 15.3\\
\hline
\multicolumn{5}{l}{{\footnotesize Assigned parameters are indicated with a \dag symbol; others are derived.}} \\
\end{tabular}
\label{t:Model_A_pars_pot}
}
\end{center}
\caption{Standard Potential Parameters of Model {\bf A}.}
\end{table}

\begin{table}
\begin{center}
{\small 
\begin{tabular}{l c c c c }
\hline
          &                & \multicolumn{3}{c}{Scaled to  $x_0 =$} \\
Parameter \& units       &    Symbol           & $1.0 \rm kpc$  &  $0.2 \rm kpc$  & $5.0 \rm kpc$ \\
\hline
\hline
\multicolumn{3}{l}{\bf Hot Atmosphere:} \\
{\dag Virial/Gas temperature} & $\beta_{\rm h}$ &{ 1.0} &{ 1.0} &{ 1.0} \\
Gas Temperature ($^\circ \rm K$)              & $T_{\rm h} $    & {$1.20\times 10^{7}$} & { $1.20\times 10^{7}$} & { $1.20\times 10^{7}$} \\
\\
\multicolumn{3}{l}{\bf Central Values:} \\
{\dag  pressure$/k$} 
($\rm cm^{-3} \> ^\circ K$)& $p_{\rm h,c}/k $ & $1.00\times 10^{6}$   & $5.00\times 10^{6}$   &  $2.00\times 10^{5}$ \\
{ pressure} 
($\rm dynes \> cm^{-2}$)& $p_{\rm h,c}$     & $1.38\times 10^{-10}$ & $6.90\times 10^{-10}$ &  $2.76\times 10^{-11}$ \\
{ number density} 
($\rm cm^{-3}$)  & $n_{\rm h,c}$          & $8.35\times 10^{-2}$  & $4.17\times 10^{-1}$  &  $1.67\times 10^{-2}$ \\
{ mass density} 
(g cm$^{-3}$)  & $\rho_{\rm h,c}$       & $8.64\times 10^{-26}$ & $4.32\times 10^{-25}$ &  $1.73\times 10^{-26}$ \\
\hline
\multicolumn{3}{l}{\bf Warm Disk--ISM :} \\
{Virial/Gas temperature}                      & $\beta_{\rm w}$ & 1200.0 & 1200.0 & 1200.0 \\
{\dag Gas Temperature}($^\circ \rm K$)              & $T_{\rm w} $    &$1.0\times 10^{4}$&$1.0\times 10^{4}$&$1.0\times 10^{4}$ \\
{\dag Turbulent dispersion}
($\rm km \> s^{-1}$)                      & $\sigma_{\rm t}$&40.0&40.0&40.0 \\
{\dag Rotational Support}                & $E_{\rm R}$&0.93&0.93&0.93 \\
\\
\multicolumn{3}{l}{\bf Internal Non--Uniformity :} \\
{\dag Log-Normal Mean }                      & $\mu $         & 1.0      & 1.0     & 1.0 \\
{\dag Log-Normal Variance}                   & $\sigma^2$ & 5.0      & 5.0     & 5.0 \\
{\dag Density power-law}                      & $\beta$       & $5/3$ & $5/3$ & $5/3$ \\
\\
{Volume of warm gas }
($\rm pc^3$) & $V_{\rm w}$                & $2.55\times 10^{7}$ & $2.04\times 10^{5}$ & $3.19\times 10^{9}$ \\
Mass of warm gas 
($M_\odot$)             & $M_{\rm w}$                 & $4.67\times 10^{5}$ & $1.87\times 10^{4}$ & $1.17\times 10^{7}$ \\
Relative Disk Mass     &                     & 1.0 & 0.04 & 25.0 \\
\\
\multicolumn{3}{l}{\bf Central Values:} \\
{ pressure$/k$} 
($\rm cm^{-3} \> ^\circ K$)& $p_{\rm w,c}/k $ & $1.00\times 10^{6}$   & $5.00\times 10^{6}$   &  $2.00\times 10^{5}$ \\
{\dag number density} 
($\rm cm^{-3}$)  & $n_{\rm w,0}$          & 10.0 & 50.0 & 2.0 \\
{ mass density} 
(g cm$^{-3}$)  & $\rho_{\rm w,0}$       & $1.04\times 10^{-23}$ & $5.18\times 10^{-23}$ &  $2.07\times 10^{-24}$ \\
\hline
\multicolumn{5}{l}{{\footnotesize Assigned parameters are indicated with a \dag symbol; others are derived.}} \\
\end{tabular}
\label{t:Model_A_pars_Gas}
}
\end{center}
\caption{Standard Hot Halo and warm disk--ISM Parameters Model {\bf A}.}
\end{table}

\section{Code and Algorithm Physics}
\label{s:code}

Our simulations use a non-relativistic Piecewise Parabolic Method (PPM) code based on code provided by J. Blondin and colleagues via the web-site 
http://wonka.physics.ncsu.edu/pub/VH-1/. We have already commented on the non-realtivistic aspects of the code in \S\ref{s:jetpars}. 

The code has been extensively reorganized for efficiency and parallel execution on the SGI Altix computer operated by the Australian Partnership for Advanced Computation. We
have also added subroutines to advect passive scalars, which track the evolution of different gases and to update the energy density, using an implicit method, when optically thin radiative cooling operates. A cooling function has been implemented which is based upon output from the MAPPINGS shock and photoionization code \citep{sutherland93c,sutherland03b}.   The cooling treatment has been extended for the present models by computing an X-ray spectrum for each temperature point in the thermal cooling function, and using the spectra to construct hard and soft X-ray maps of the thermal gas, as the simulation proceeds. This is motivated by the need to improve the correspondence between simulation and observation, and to calculate more directly observable output variables, compared with the less amenable hydrodynamical variables such as density or pressure.

We have also added code to deal with numerical instabilities that can occur in strong shocks, especially when cooling is present \citep{sutherland03a}.  This code is simply referred to as \emph{ppmlr} for Piecewise Parabolic Method Lagrangian Remap.   An important advantage of this PPM algorithm is the relatively low diffusions it exhibits, compared to other finite volume method for example, allowing the number or cells needed to capture shock structures to be minimized, important in 3D simulations, and the speed of the algorithm allows the calculation of uniformly high resolutions where more complex algorithms, such as a full MHD treatment may not be practical.

 The  neglect of magnetic fields is justified in the first instance because of the large increase in simulation phase space that this entails. For example, some simulations incorporate an initially toroidal, unidirectional jet magnetic field; others inject a random field.The negelect of magnetic field probably does not have a major effect on the evolution of the simulation for the following reasons. Much of the flow in these simulations is turbulent and a completely disordered magnetic field behaves as a polytropic gas with a ratio of specific heats, $\gamma = 4/3$. This is not too different from the $\gamma$ of $5/3$ that we use here so that we expect a turbulent magnetic field to track the gas pressure reasonably well. Of course there are qualifications to this. Magnetic fields and ideal gases behave differently in shocks and a systematic component of magnetic field may introduce different dynamics than an isotropic turbulent field. Moreover, in order to investigate important Faraday rotation and polarisation effects as well as the details of the nonthermal emissivity, it is essential to include a magnetic field. These effects  are ones that we can investigate with MHD simulations. Nevertheless, for now, the current simulations provide us with a base for future work in which magnetic fields are included in a systematic way. Moreover, the general features of the flow -- obstruction by clouds, the initial formation of bubbles and the formation of radiative shocks are probably well captured by the simulations that we present.
 
We have implemented the following boundary conditions in these simulations: For most of the left hand boundary plane ($x=0$) the boundary is reflecting. The exception is the jet inlet $y_{\rm c}^2 + z_{\rm c}^2 < r_{\rm jet}^2$ where $y_{\rm c}$ and $z_{\rm c}$ are the zone-centered $y$ and $z$ coordinates of a cell and $r_{\rm jet}$ is the jet radius. This inlet region has pure inflow boundary conditions. On all other boundaries inflow/outflow conditions were used.

\section{ Results }
\label{s:results}

In this section we present the results of Model~{\bf A} in the form of multi-panel snapshots from significant epochs; these montages are designed to bring out the relevant physics of the simulations. Some snapshots correspond to slices of important dynamical variables such as density and pressure; in some cases we also present projected versions of variables such as the density. We also present volumetric ray-traced projected images of the radio and X-ray emissivity to produce synthetic images of radio and X-ray surface brightness. 

\subsection{Evolutionary phases}

In all cases the times chosen for the sequence of snapshots corresponds to the following phases in the simulation.  The mid-plane density slices (Figure~\ref{f:densslice}) most clearly illustrate the phases enumerated here, with the other representations highlighting some specific facets.
\begin{enumerate}

\item Flood and Channel phase: Snapshots at 5, 10, 15 and 25 kyr represent the time over which the jet is making its way through the porous, fractal disk.  The shape of the interacting region is amorphous, and determined by flow of hot high pressure gas along weaker line in the dense disk medium.  The pressure in this phase is very high, and the X-ray emission is a strong function of time, depending on a combination of the amount of disk material that is advected and the amount of energy that is processed by radiative shocks, which increases with the size of the region.  It is not not well determined what the time dependence of thermal luminosity is in this phase, but it is definitely super-linear.   

In the next two phases an energy-driven bubble forms and evolves, although the beginning and end of this evolution exhibit some different characteristics.
  
\item Energy-driven bubble phase: Snapshots at 35, 45 and 55 kyr represent the epoch during a high pressure, pseudo-spherical bubble forms and grows larger than the disk. The jet is still disrupted by disk material, some of which has been advected to higher altitudes, but the bulk of the energy flux drives the expansion of the nearly adiabatic bubble. A corresponding drop in the efficiency of conversion of jet energy to thermal emission is seen at about 20~kyr. Most of the bubble grows (outside the disk) with a power-law that is consistent with classical energy bubble theory. The flood and channel behavior persists somewhat within the plane of the disk, but this involves an ever decreasing fraction of the jet energy flux.

\item Jet-breakout phase: In the epochs represented by 55, 65, and 70 kyr the jet starts to break fee of the few remaining clouds in its path and the jet terminus starts to propagate towards the edge of the bubble. The bubble has a generally low density, so that once clear of dense material and re-collimated, the jet transits the bubble quickly.

\item Classical phase: At 75, 85, and 95 kyr and beyond the jet pierces the original bubble and then starts to form a classical radio lobe with hotspot, cocoon, bow-shock and back flow.  The remnant of the spherical bubble continues to grow, and flow continues within the disk, with many radiative and non-radiative shocks throughout the whole disk persisting to late times in all locations bar the very centre cleared by the main jet.

\end{enumerate}

\clearpage

\subsection{Density}
\label{s:density}

\subsubsection{Mid-plane density slices}

\begin{figure}
\begin{center}
\includegraphics[width=5in]{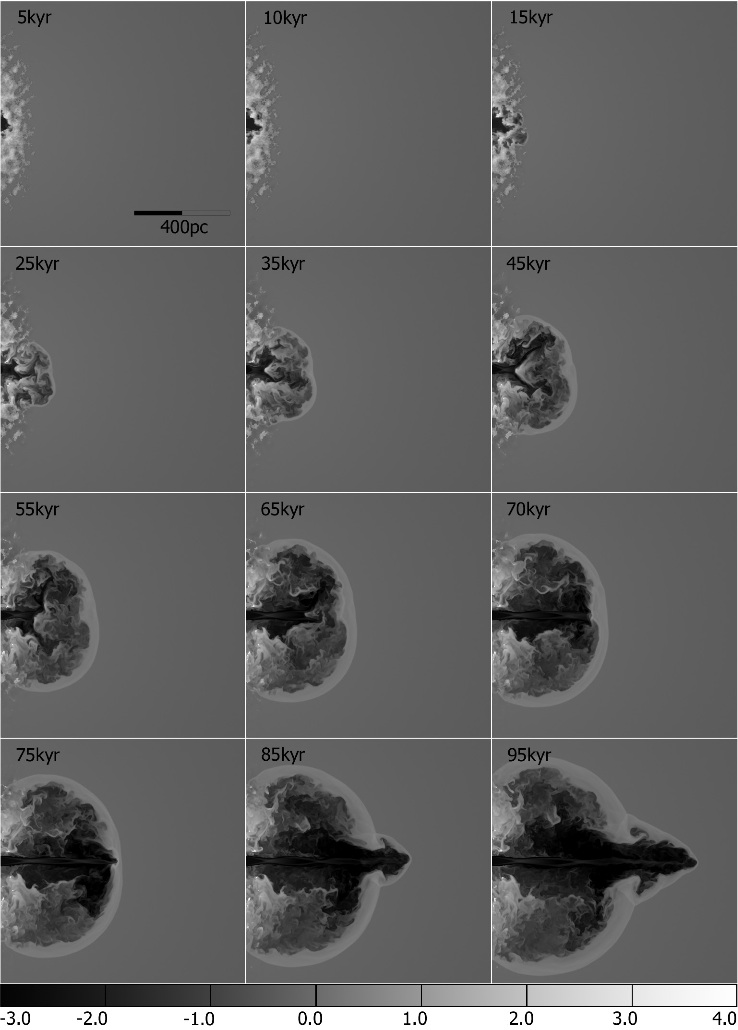}
\caption{\footnotesize Mid-plane density slices: The panels represent the logarithm of the density in a $k=255$ plane of the $512^3$ simulation at the various phases of the evolution described at the beginning of \S\ref{s:density}. The grayscale bar shows the range of the density. The panels from 5 -- 25~kyr are typical of the flood and channel phase; the panels from 35 -- 55~kyr are representative of the energy-driven bubble phase; the jet-breakout phase extends from approximately 55 -- 70~kyr and the classical phase extends from 75 -- 95 kyr and beyond.  The selection of snapshot times is the same in all subsequent figures.}
\label{f:densslice}
\end{center}
\end{figure}

\begin{figure}
\begin{center}
\includegraphics[width=5in]{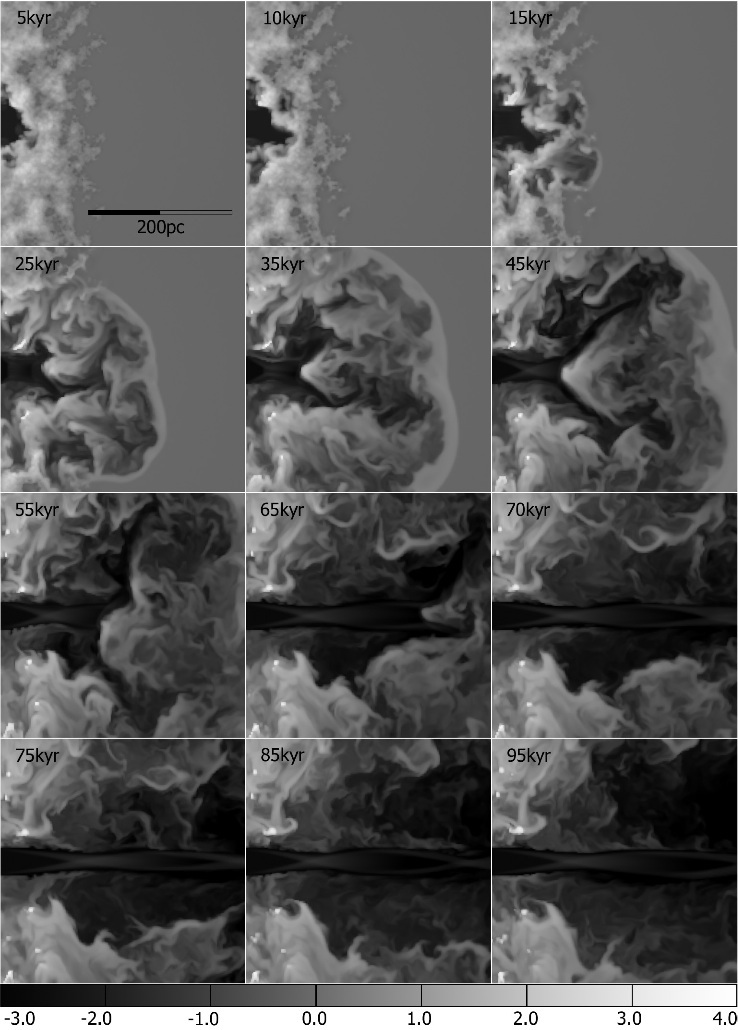}
\caption{\footnotesize Mid-plane density slices: The panels represent the logarithm of the density in a $170 \times 170$ cell region of the $k=255$ plane of the $512^3$ simulation. The galaxy core is at the center of the left hand edge. The grayscale bar shows the range of the density.}
\label{f:innerdensslice}
\end{center}
\end{figure}

One of the best ways to initially gauge the features of a three dimensional simulation is via density slices along the mid-plane of the simulation. In Figure~\ref{f:densslice} we show such a sequence of images covering the entire computational grid. In Figure~\ref{f:innerdensslice} we show a zoomed in view which better illustrates the features in the density close the nucleus, emphasizing the high resolution of this simulation. 

In the initial flood and channel phase the jet starts to clear a path through the porous disk and begins to break out. The corresponding panels in Figure~\ref{f:innerdensslice} show this initial phase of the high resolution simulation in more detail. At 10 and 15 kyr in particular, there are ``bright'' spots of high density caused by radiative shocks driven into the warm gas. The jet is anything but straight as it selects a path of least resistance through the region in which the initial density is  described in terms of a log-normal distribution (see \S~\ref{s:lognormal}).

In the energy-driven bubble phase (35 -- 55 kyr), the jet starts to clear a path through the disk. However, the jet energy and momentum are still spread over a wide solid angle. A dense obstruction causes the jet to split (at least in the mid-plane). Further regions of high density are observed, often at the tips of dense cloud material that is being ablated by the outflowing radio plasma. The thermalized jet pressure drives an almost spherical bubble in the surrounding medium. A bow-shock surrounds the bubble and the contact discontinuity between the bubble and the hot interstellar medium is apparent.

In the jet-breakout phase (55 -- 70~kyr) the pressure of shocked jet plasma, continues to drive a more nearly spherical bubble into the hot interstellar medium. Towards the end of this phase, the obstructing cloud that was responsible for the remaining disruption of the jet dissipates and the jet rapidly crosses the bubble. 

At 75~kyr (the lowest left panel of Figure~\ref{f:densslice}) a collimated jet flow has been established and the jet is about to pierce the initial bubble. Subsequently, the jet propagates beyond the bubble and starts to establish a classic double lobe morphology consisting of bow-shock, cocoon  and backflow. 
  
Another way of looking at this phase is that the quasi-isotropic (energy-driven) phase of radio lobe expansion makes a transition to a bow-shock dominated (momentum-driven) phase as the now relatively undeflected jet progresses through the interstellar medium. An interesting feature, which is discussed further below, is that the jet, whilst maintaining a more or less single direction, is unstable. As a result of both filamentation and helical instabilities, the end of the jet has become somewhat diffuse and the jet momentum is spread over a region which is wider than the original jet diameter in a similar manner to that originally envisaged by \citet{scheuer82a}.

\subsubsection{Face-on column density}

\begin{figure}
\begin{center}
\includegraphics[width=5in]{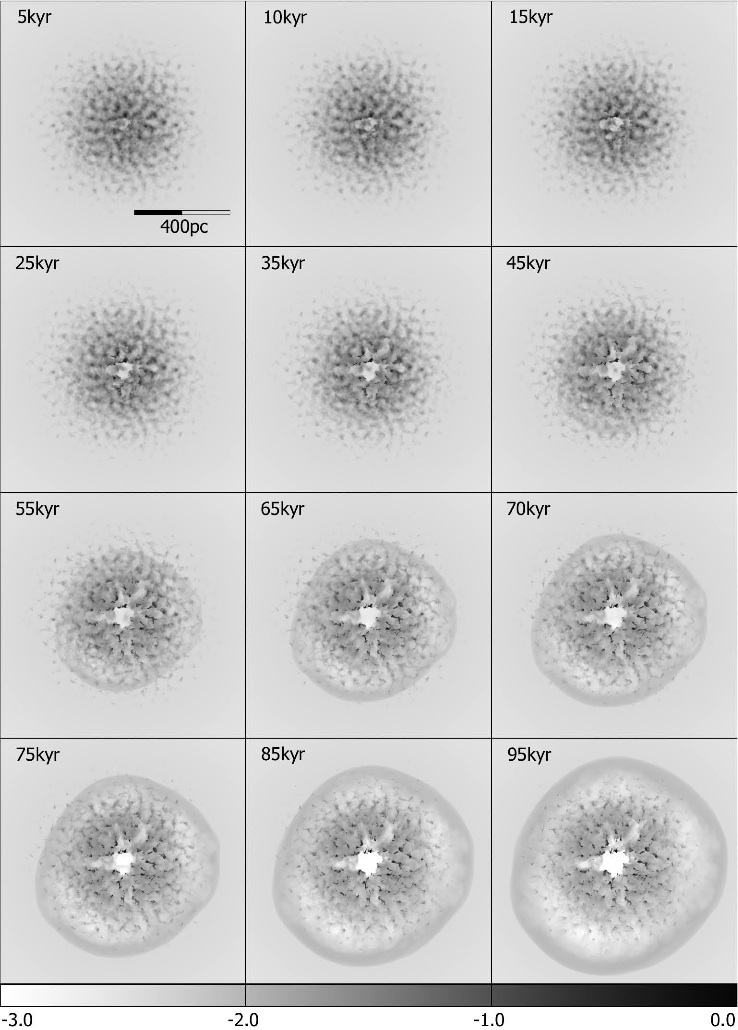}
\caption{\footnotesize Face-on column density: The panels represent the logarithm of the column density viewed along the jet direction.}
\label{f:dens_column}
\end{center}
\end{figure}

Figure~\ref{f:dens_column} shows a different view of the evolution of the density provided by images of the \emph{column} density, which are obtained by projection in the direction of propagation of the jet. As well as serving to provide a more complete picture of the evolution, these images, taken in conjunction with the previous density slices also show that the jet -- ISM interaction has real three dimensional characteristics.

From 5 -- 15~kyr the jet--disk interaction is obviously confined to the region of the disk adjacent to the jet orifice. From 25 -- 45~kyr channels of low density plasma are formed in the disk corresponding to the path of least resistance in the fractal density distribution to the dynamic pressure of the jet plasma. At the same time numerous high density regions appear (the whitest regions in the grayscale images) indicating the locations of radiative shocks in which the density has increased by up to a factor of 100.

During the 55 -- 70~kyr phase (the jet breakout phase described above) the bubble of low density gas starts to extend beyond the boundary of the disk. By 70~kyr a ring of shocked hot ISM is apparent, defining the base of the spherical bubble. The channels produced during the previous phases become more clearly defined. 

There is little qualitative change in the appearance of this projection save for the fact that the base of the bubble grows, consistent with the growth seen in the slice images. However, the fact that there is no major change in the appearance of the fractal disk is a feature of interest. This signals that radiative shocks continue to be driven into the dense gas as a result of the pressure in the bubble. We discuss this further in \S\ref{s:xray}  when discussing the X-ray emission.
 
\subsection {Pressure}
\label{s:pressure}

\begin{figure}
\begin{center}
\includegraphics[width=5in]{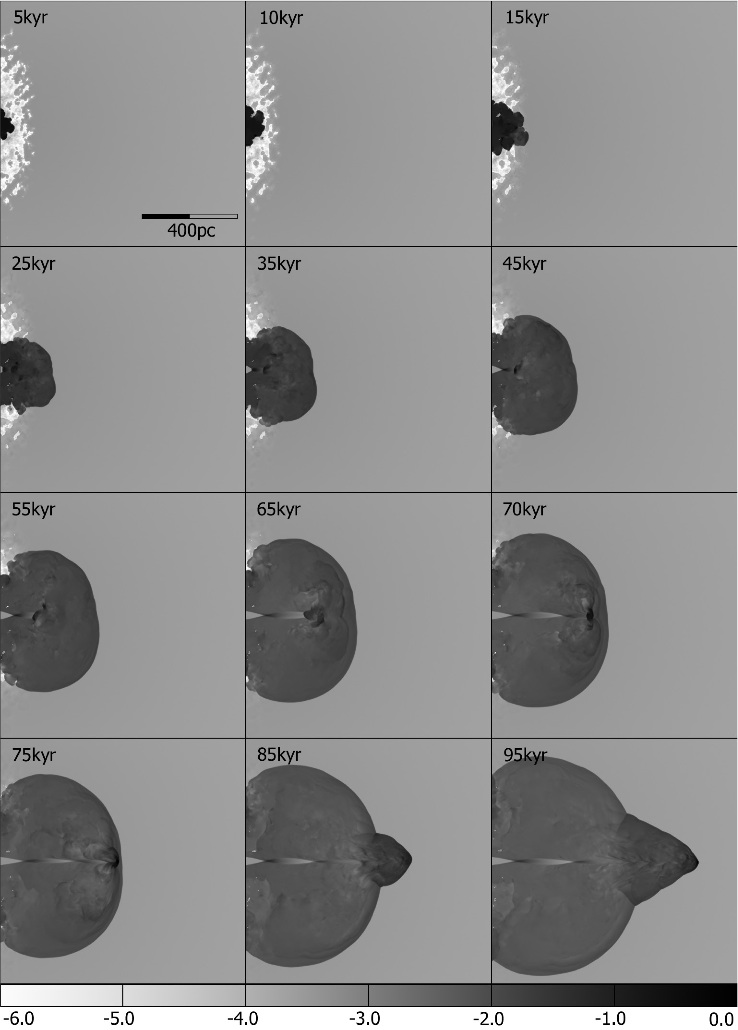}
\caption{\footnotesize Mid-plane pressure slices: The panels represent the logarithm of the pressure in an $k=255$ plane of the $512^3$ simulation at the same phases of the evolution as in Figure~\ref{f:densslice}. Note that for clarity, the sense of the grayscale colorbar has been reversed compared to that for the density.} 
\label{f:presslice}
\end{center}
\end{figure}

Mid-plane pressure slices in Figure~\ref{f:presslice} confirm the evolutionary features revealed by the density and at the same time show some additional characteristics. 

The flood and channel phase (5 -- 25~kyr) shows a high-pressured amorphous region where the jet struggles to make its way out of the confining disk. The lower pressure dense gas (the white region) surrounds this amorphous bubble.  

In the energy-driven bubble phase (35 -- 55~kyr) the pressure in the jet-driven bubble is fairly uniform (as expected) but also shows at 35 kyr a biconical shock associated with the compression of the emerging jet by the overpressured bubble. and a bow shock caused by obstructing high density gas in the path of the jet. There are also some spots of high pressure associated with the general turbulence in the bubble. 

During the jet-breakout phase (55 -- 70~kyr) most of the radio lobe is at constant pressure, and there are only small residual regions of low pressure associated with the disk. Further biconical shocks form within the jet and the high pressure jet terminus is seen propagating through the bubble. 

During the classical phase (75 -- 95~kyr the hot spot at the end of the jet pierces the bubble and forms a higher pressure, momentum-driven cocoon. The biconical shocks do not persist within this cocoon and the high pressure region at the end of the jet is spread out over more than a jet diameter. This is a result of the jet jittering noted in \S\ref{s:density}.

\clearpage

\subsection{Radio Surface Brightness}
\label{s:inu}

\subsubsection{Formation of synthetic radio images}

We have constructed a surface brightness emissivity from the pressure $P_{\rm nt}$ of non-thermal plasma, defined to be the material which originates from the jet. The emissivity is defined by:
\begin{equation}
j_\nu \propto P_{\rm nt}^{(3+\alpha)/2} \, \nu^{-\alpha} 
\label{e:emissivity}
\end{equation}
where $\alpha$ is the spectral index. This form of the emissivity assumes that the magnetic pressure tracks the plasma pressure, that is, $B^2/8 \pi \propto P$.  The scalar tracer $\phi$, which is the mass concentration by density of the non-thermal plasma, is used to indicate the presence of jet plasma at the various regions of the grid. An approach that we have used in previous papers (e.g. \citet{saxton05a}) is to simply take the emissivity proportional to the tracer. This has the desired effect of making the emissivity non-zero where $\phi=1$ (the jet value) and zero in the background, where $\phi=0$. This ansatz is innapropriate in regions of intermediate values of $\phi$ since the nonthermal pressure, which determines the jet emissivity, dominates even when the ratio of non--thermal to total density is small. Hence, in order to gain a more accurate idea of the distribution of non-thermal emission we use the above expression (\ref{e:emissivity}) for the emissivity when $\phi$ is greater than some threshold value, usually about $10^{-3}$.

The radio images produced from this emissivity do not include the effects of aberration, relativisitic beaming, or spectral steepening.  They correspond best to low frequency observations where electron-aging effects are less.  They are intended as reasonable indicators of the distribution of radio emission.  The detailed modelling of all the possible non-thermal emission processes (primarily synchrotron and inverse Compton) from a frame of the simulation to the level required for a point by point prediction and comparison with a specific radio observation is beyond the scope of the current investigation.  In future, with a fuller relativistic and MHD treatment, detailed radio emission predictions will be included.

The radio thermal emissivity cubes are ray-traced using a volumetric ray-tracing program, {\em render},  to provide synthetic surface brightness images, as well as for column densities and other more general projection of variables.   {\em render} is a 2D and 3D volumetric renderer that preserves surface brightnesses, and uses homogeneous coordinates for projecting a range of cartesian and other geometries, under a range of transformations, including perspective (not utilized here), plus rotations, affine transforms and stereoscopic modes.  

The {\em render} code also handles a range of geometries and symmetries, and in the double jet renderings of the radio emission (e.g. in Figures~\ref{f:inu_A} and \ref{f:inu_B}) the simulation data are reflected in  the $yz$ plane, perpendicular to the initial jet direction during the ray-tracing process.  Perspective and other depth cueing transformations have been turned off for all the following figures, resulting in the projection as seen from an infinite distance from the source.  

\subsubsection{Images of Model {\bf A}}

\begin{figure}
\begin{center}
\includegraphics[width=5in]{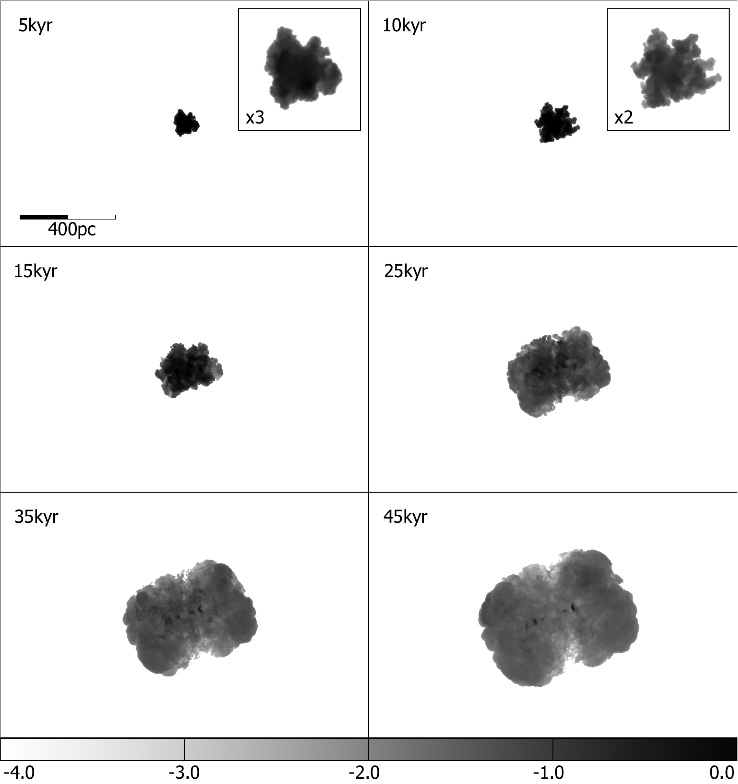}
\caption{\footnotesize Radio Surface Brightness: The panels represent the logarithm of the synthetic radio  surface brightness at the 5 -- 45~kyr phases of the evolution, corresponding to the first 6 panels of the density evolution in Figure~\ref{f:densslice}. The insets for 5~kyr and 10~kyr show the surface brightness images magnified by factors of 3 and 2 respectively in order to bring out the detail in these early stages.}
\label{f:inu_A}
\end{center}
\end{figure}

\begin{figure}
\begin{center}
\includegraphics[width=5in]{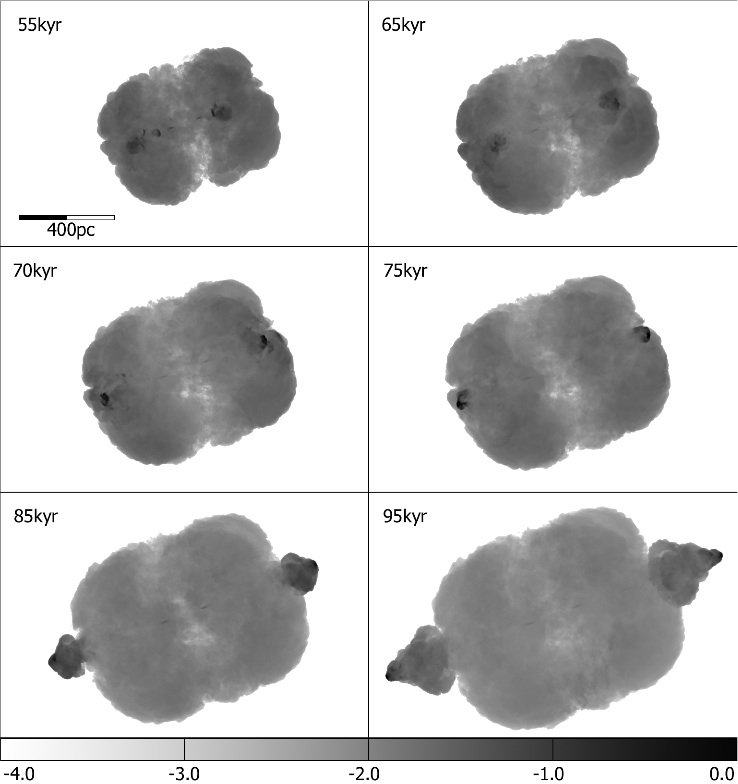}
\caption{\footnotesize Radio Surface Brightness: The panels represent the logarithm of the synthetic radio  surface brightness at the 55 -- 95~kyr phases of the evolution, as in the second 6 panels of Figure~\ref{f:densslice}.}
\label{f:inu_B}
\end{center}
\end{figure}

A series of synthetic radio surface brightness snapshots is displayed in Figures~\ref{f:inu_A} and \ref{f:inu_B} at the same times as in the previous figures of desnity and pressure slices. Note that the panels are split across two figures in order not to lose detail in the double-sided surface brightness images.

At the early times (5 -- 15~kyr) the surface brightness shows the amorphous structure that is evident in the density slices discussed above.

In the flood and channel phase from approximately 25 -- 45~kyr, we see the formation of the two radio-emitting bubbles with some flocculent structure evident in the surface brightness resulting from the non-uniform thermalisation of jet plasma during this phase. We also see quite clearly (particularly at 35~kyr and 45~kyr) the bow shock (on each side) caused by the impact of the jet on the obstructing cloud revealed in the density and pressure slices discussed above.

During the jet-breakout phase from 55 -- 70~kyr, the radio-emitting bubbles become smoother in appearance but high surface brightness features associated initially with the jet-cloud bow shock (55~kyr) and then with the jet terminal shock (65 and 70~kyr) are also seen propagating out through the respective bubbles. Another feature which becomes obvious here, but which is also evident from approximately 25~kyr onwards, is a band of low surface brightness emission caused by the partial exclusion of radio-emitting plasma from the disk region. 


From 75~kyr onwards we see the effect of the jets breaking free of the bubbles.

\clearpage

\subsection{X-ray emission}
\label{s:xray}

\subsubsection{Formation of images}

Below, we show X-ray images calculated for specific X-ray emission bands. These images are calculated as follows. First, the X-ray spectrum, corresponding to the cooling function used in the calculations is calculated as a function of temperature. This is a by-product of the original evaluation of the cooling function. In the range 0.1--10keV the spectra consist of 1100 energy bins. Second, integrals over specified sub-bands are evluated and the fraction of the total cooling in those bands is evaluated. Third, the fractions of the total cooling in each band are tabulated. The {\em render} code then uses the total emissivity and temperature cubes to determine the volume emissivity in each band of interest and then ray-traces the emissivity to form the images presented below in \S\ref{s:xray.images}.

Self-absorption is neglected, as is any form of scattering, thus restricting this analysis of the thermal emission to the X-ray domain for the present. The softest X-ray synthetic images are approximate since the neglected self-absorption is stronger for energies $\sim 100\> \rm  eV$.  Despite this limitation the synthetic X-ray images give a good indication of the locations where soft or hard X-rays should dominate, although we don't assign precise absolute values to the X-ray surface brightnesses at present, leaving them as relative fluxes. Three color X-ray images as well as single band images can be generated but these are not presented here.

\subsubsection{Description of images from Model {\bf A}}
\label{s:xray.images}

\begin{figure}
\begin{center}
\includegraphics[width=5in]{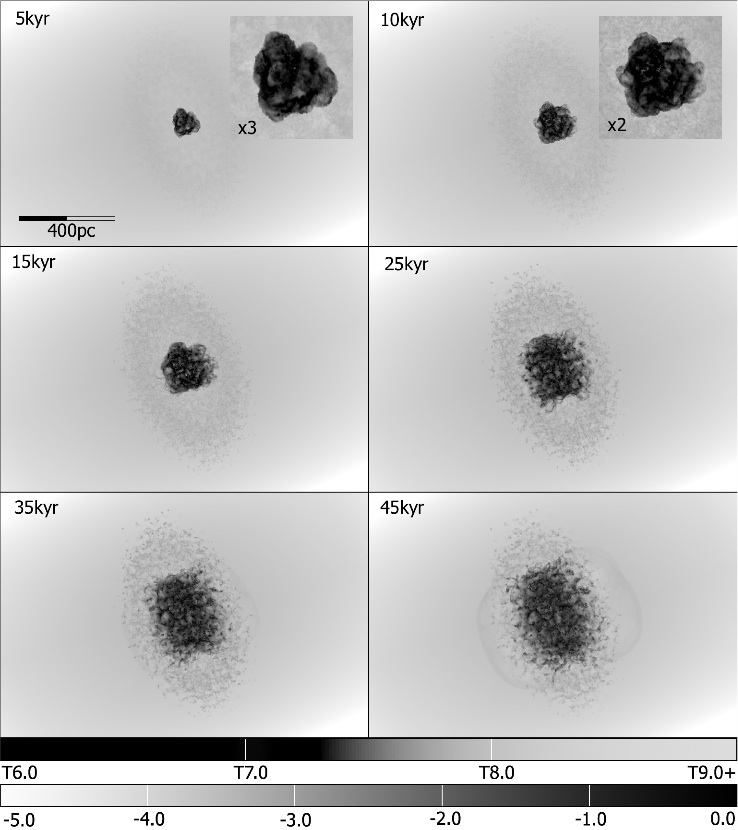}
\caption{\footnotesize  Soft X-rays: Snapshots of the logarithm of the soft X-ray surface brightness between 5 and 45 kyr. The lower grayscale bar indciates the scale of surface brightness in the image. The upper grayscale bar with numbers prefixed by `T' indicates the relative contribution of gas at various temperatures to the emissivity. Dark indicates a large contribution; light indicates a low contribution. }
\label{f:xsoft1}
\end{center}
\end{figure}

Snapshots of the soft (0.1 -- 0.75~keV) and hard (1.5 -- 10~keV) X-ray surface brightness are presented in Figures~\ref{f:xsoft1}, \ref{f:xsoft2}, \ref{f:xhard1} and \ref{f:xhard2}.

In order to indicate the contribution from gas at various temperatures to the emission in the soft and hard bands, we have included a grayscale bar in each X-ray image which indicates the relative contribution of gas at a specific temperature to the emissivity in the specific band. Light indicates a low relative contribution; dark indicates a high relative contribution. Thus the dominant contribution in the soft X-ray band is from gas in the range $10^6 \> {\rm K}\la T \la 2 \times 10^7 \> \rm K$. However, there is also some contribution from gas at higher temperatures up to and exceeding $10^9 \> \rm K$. In the hard X-ray band there is little contribution from gas whose temperature is below $10^7 \> \rm K$; the dominant contribution is from gas whose temperature exceeds approximately $2 \times 10^7 \> \rm K$.

\subsubsection{Soft X-rays}

During the period from 5--45~kyr the soft X-ray emission is dominated by emission from the disk as the developing bubble of non-thermal plasma sweeps through driving moderate velocity radiative shocks into the dense gas concentrations within the log-normal disk. The bright (i.e. dark color) emission is the result of the slower shocks producing a temperature $\sim \hbox{a few} \times 10^6 \> \rm K$ in the high density material. The brightness of the emission is also enhanced by the density squared dependence of the emissivity.

Some faint emission from the gas at the outer parts of the disk is also visible resulting from the driving of radiative shocks into the disk by the higher pressure hot interstellar medium. This is a consequence of the way in which the initial distribution of gas has been established and not too much should be read into the details of this emission. 

\begin{figure}
\begin{center}
\includegraphics[width=5in]{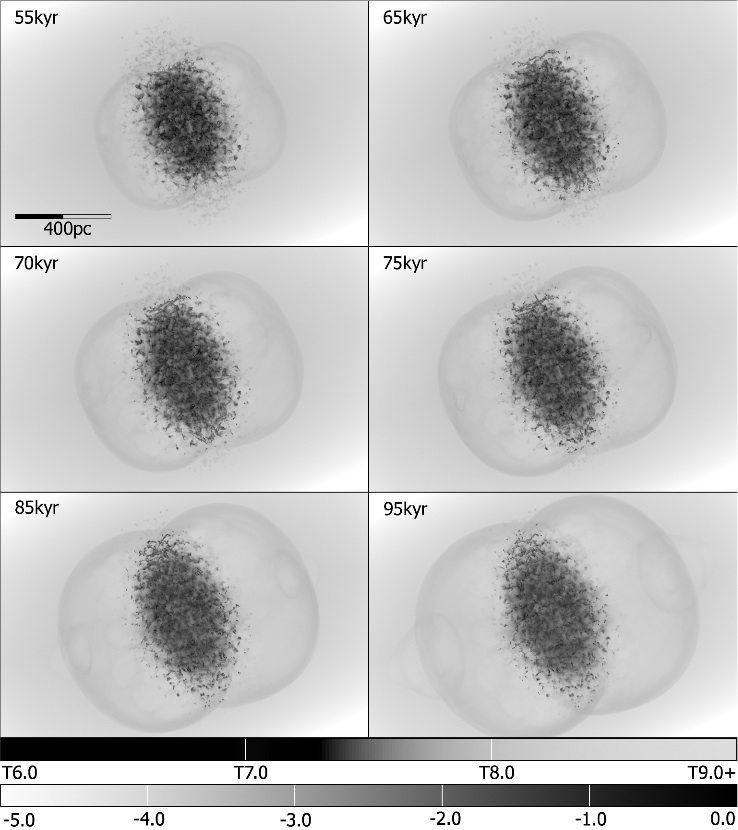}
\caption{\footnotesize Soft X-rays: Snapshots of the logarithm of the soft X-ray surface brightness between 55 and 95 kyr. The upper grayscale bar with numbers prefixed by `T' indicates the relative contribution of gas at various temperatures to the emissivity. Dark indicates a large contribution; light indicates a low contribution.}
\label{f:xsoft2}
\end{center}
\end{figure}

As the simulation evolves (Figure~\ref{f:xsoft2}) an additional feature appears -- X-ray emission from the spherical bubbles blown by the jets on either side of the double source. At 95~kyr, faint features corresponding to the piercing of the bubbles by the jets are just perceptible.

\begin{figure}
\begin{center}
\includegraphics[width=5in]{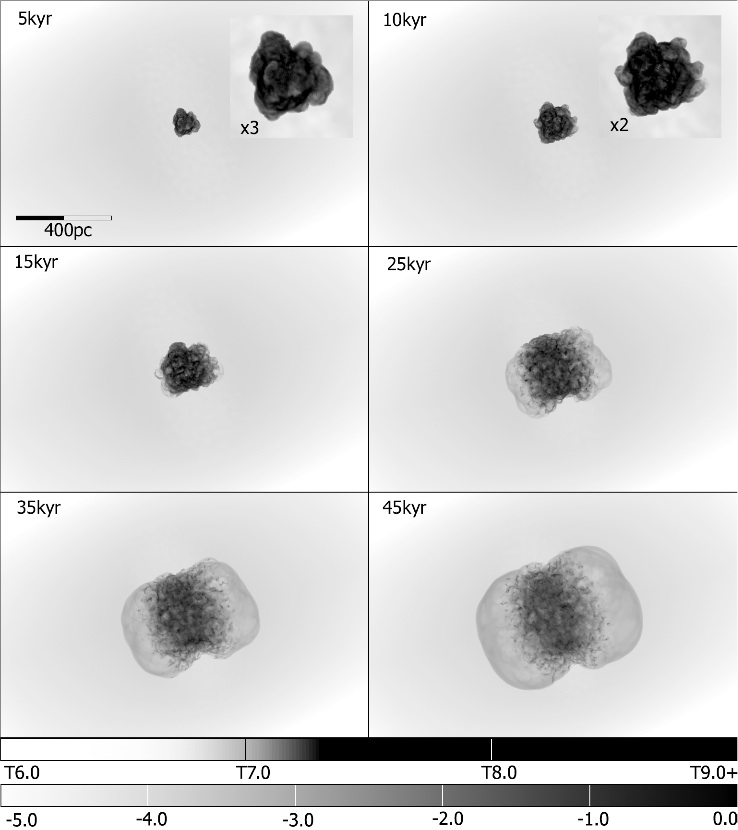}
\caption{\footnotesize Hard X-rays: Snapshots of the logarithm of the hard X-ray surface brightness between 5 and 45 kyr. The upper grayscale bar with numbers prefixed by `T' indicates the relative contribution of gas at various temperatures to the emissivity. Dark indicates a large contribution; light indicates a low contribution.}
\label{f:xhard1}
\end{center}
\end{figure}

\begin{figure}
\begin{center}
\includegraphics[width=5in]{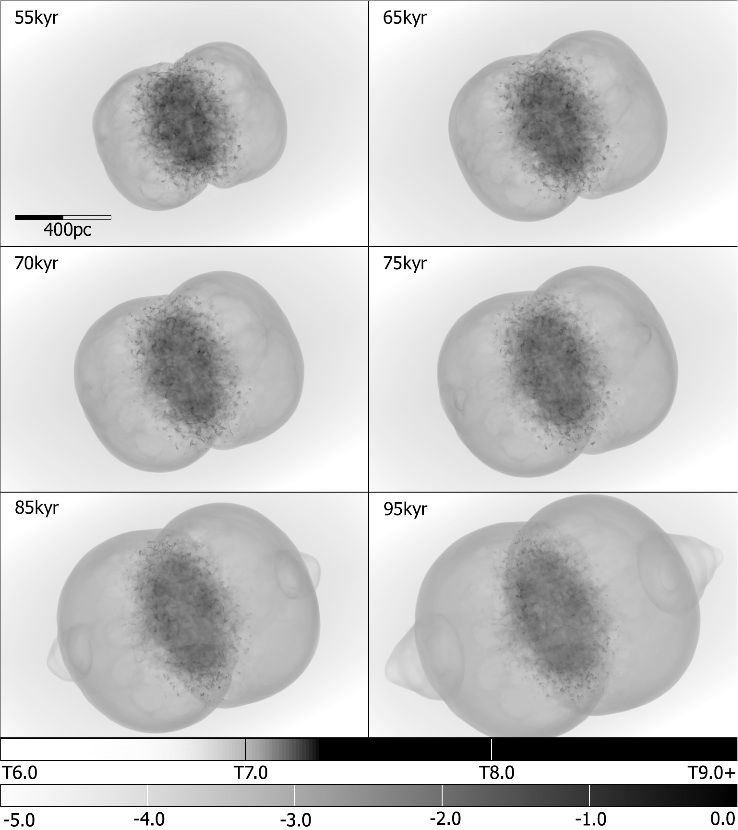}
\caption{\footnotesize Hard X-rays: Snapshots of the logarithm of the hard X-ray surface brightness between 55 and 95 kyr. The upper grayscale bar with numbers prefixed by `T' indicates the relative contribution of gas at various temperatures to the emissivity. Dark indicates a large contribution; light indicates a low contribution.}
\label{f:xhard2}
\end{center}
\end{figure}

The hard X-ray emission (Figures~\ref{f:xhard1} and \ref{f:xhard2}) presents a similar evolutionary picture with some interesting differences. The first obvious feature is that the X-ray emission from the jet-driven bubbles is much brighter relative to the disk than in the soft X-ray case and the ``dimples'' on the edge of the bubbles caused by the jet breakout and the formation of the classical radio lobe structure are more apparent both at 85 and 95~kyr. The simple reason for this is that the shocks associated with the bubbles are much faster than those driven into the disk and shock the ambient gas to a much higher temperature -- typically $\gg 10^9 \> \rm K$ compared to $\sim \hbox{a few} \times 10^6 \> \rm K$. 

The second feature is that the hard X-ray emission from the disk is smoother than the soft X-ray emission.. This is the result of the emission from faster shocks being driven into lower density gas in between the dense concentrations within the disk. 

\begin{figure}
\begin{center}
\includegraphics[width=5in]{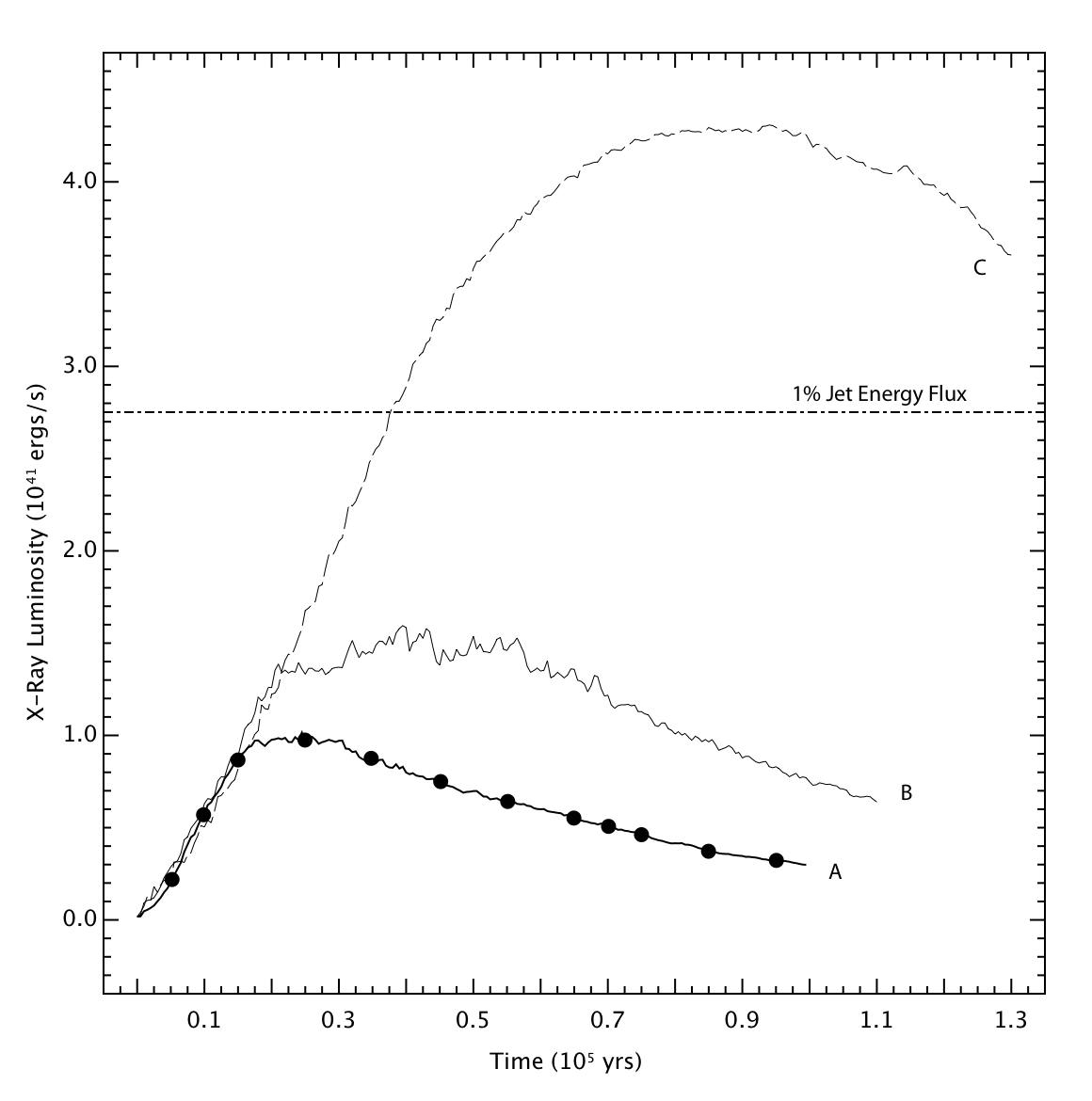}
\caption{\footnotesize The evolution of the $0.1-10.0 \> \rm keV$ X-ray luminosity of models {\bf A} , {\bf B}  and {\bf C}. The times of the snapshots used in the detailed description of model {\bf A} are indicated by filled circles. A line indicating 1\% of the jet energy flux is also drawn.}
\label{f:L_X}
\end{center}
\end{figure}

One of the most significant features of both the hard and soft X-ray emission  is that the disk remains bright well after the jet has broken free and started to form a radio lobe. This occurs because the high-pressured bubble continues to drive shocks into the disk gas. However, the brightest phase of the radio plasma -- jet interaction occurs when the jet plasma is forcing its way through the disk. This is evident from the images and more quantitatively from the plots of X-ray luminosity (Figure~\ref{f:L_X}), which we discuss further below.

\begin{figure}
\begin{center}
\includegraphics[width=5in]{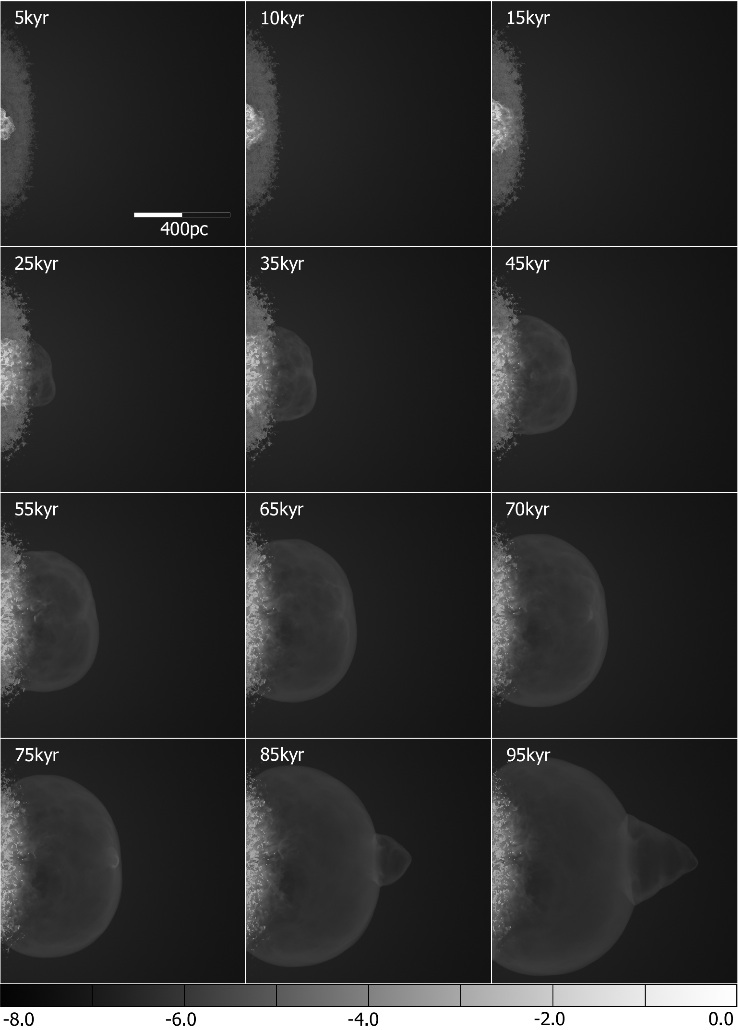}
\caption{\footnotesize Integrated Thermal Emissivity: The panels represent the logarithm of the cooling over the entire grid. }
\label{f:totcool}
\end{center}
\end{figure}

\begin{figure}
\begin{center}
\includegraphics[width=5in]{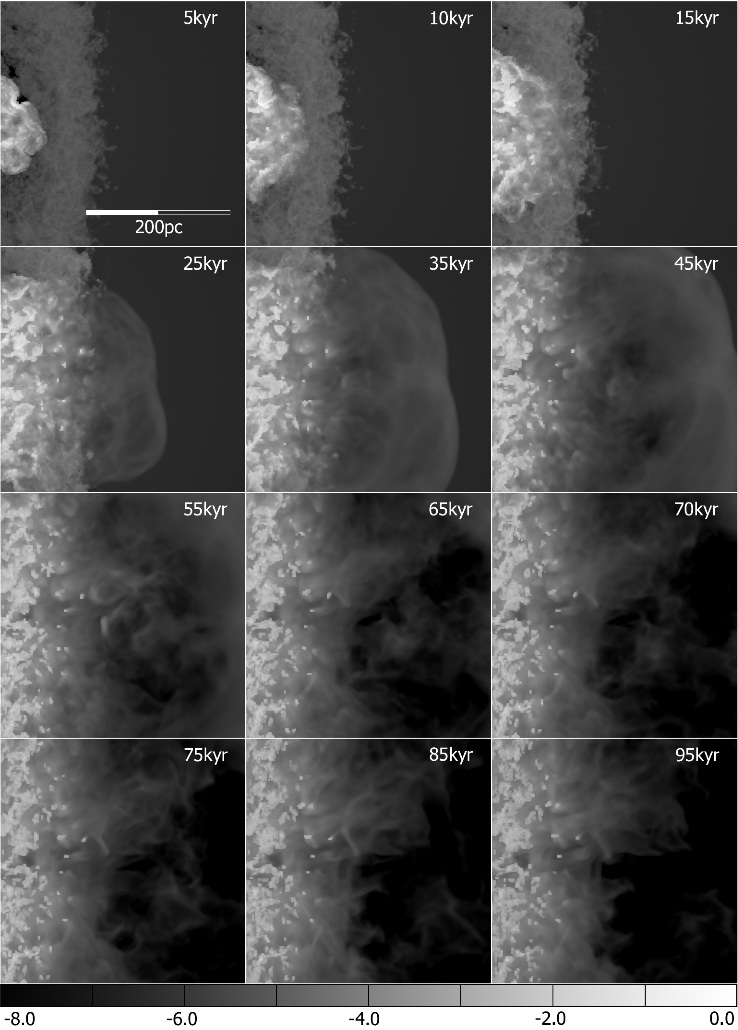}
\caption{\footnotesize Integrated Thermal Emissivity: The panels represent the logarithm of the cooling in the inner $170^3$ cell region, the same region sliced in figure 3. The grayscale bar shows the range of the total cooling.}
\label{f:innertotcool}
\end{center}
\end{figure}

In addition to the general X-ray luminous features discussed above, we can see  wisps of radiating gas, which are entrained into the lower regions of the bubble. Figures \ref{f:totcool} and \ref{f:innertotcool} show the overall contribution to total thermal cooling  dominated by the clumpy disk material, and show how some of the entrained ISM gas forms filaments which cool and increase in contrast at late times, as well as the great brightness of the dense cloud cores that remain in the disk throughout. Filaments are particularly evident in the jet breakout phase from 55 -- 70~kyr.  This entrainment is also discussed further below in \S\ref{s:transport}.

The $0.1 - 1.0 \> \rm keV$ X-ray emission for models {\bf A}, {\bf B}  and {\bf C} as a function of time is shown in Figure~\ref{f:L_X}. The common feature of all the luminosity curves is that they rise approximately linearly to a broad maximum and then decline. For the more realistic models  {\bf A} and  {\bf B}, the maximum luminosity is a fraction  of one percent of the jet energy flux. In the physically unrealistic model~ {\bf C}, the peak X-ray luminosity is about 1.6\% of the jet energy flux.    In all cases, the peak X-ray luminosity occurs when the jet can be considered to have just broken free of the disk. Before that epoch a portion of the jet power is being directed into the radiative shocks that are responsible for the X-ray emission. When the jet breaks out, the fraction of its power that was being diverted into radiative shocks is reduced since the jet power is now diverted into non-radiative shocks in the hot interstellar medium.

Given the major morphological disruptions revealed by these simulations (especially for radio plasma), why is the efficiency of conversion $\sim 1 \%$ of jet power into X-ray luminosity so low?
This is mainly the result of the relative density of the jet and clouds: Let the density of the jet be $\rho_{\rm jet}$ and the density of a cloud be $\rho_{\rm cl}$. Then, momentum balance in the jet--cloud shock system gives us the well-known result that the velocity of the radiative cloud shock is $v_{\rm sh} \approx (\rho_{\rm jet}/\rho_{\rm cl})^{1/2} v_{\rm jet}$. The energy flux density in the shocked gas, which is converted into radiation is given by $1/2 \rho_{\rm cl} v_{\rm sh}^3 \approx 1/2 (\rho_{\rm jet}/\rho_{\rm cl})^{1/2} \rho_{\rm jet} v_{\rm jet}^3$. Hence the efficiency of conversion of jet power into radiation is partly determined by the factor $(\rho_{\rm jet}/\rho_{\rm cl})^{1/2}$ which is of order $10^{-2}$ here. 
Given the efficiency factor $\sim 1\%$ associated with the jet to cloud density ratio it is therefore unsurprising to find that the peak efficiency is of order 1\%.  

The other factor affecting the conversion of jet power to radiation is the effective cross sectional area of the jet that is tapped to produce shocks. This depends, at any one time, upon how much of the jet is obstructed by dense gas. Hence, we see an almost identical rise in X-ray luminosity in models  {\bf A},  {\bf B} and  {\bf C} as the jet intersects more gas as it attempts to break free and the increasing peak luminosity from Model {\bf A} to {\bf C}, reflecting the longer period of obstruction as we progress through the models.

\begin{figure}
\begin{center}
\includegraphics[width=5in]{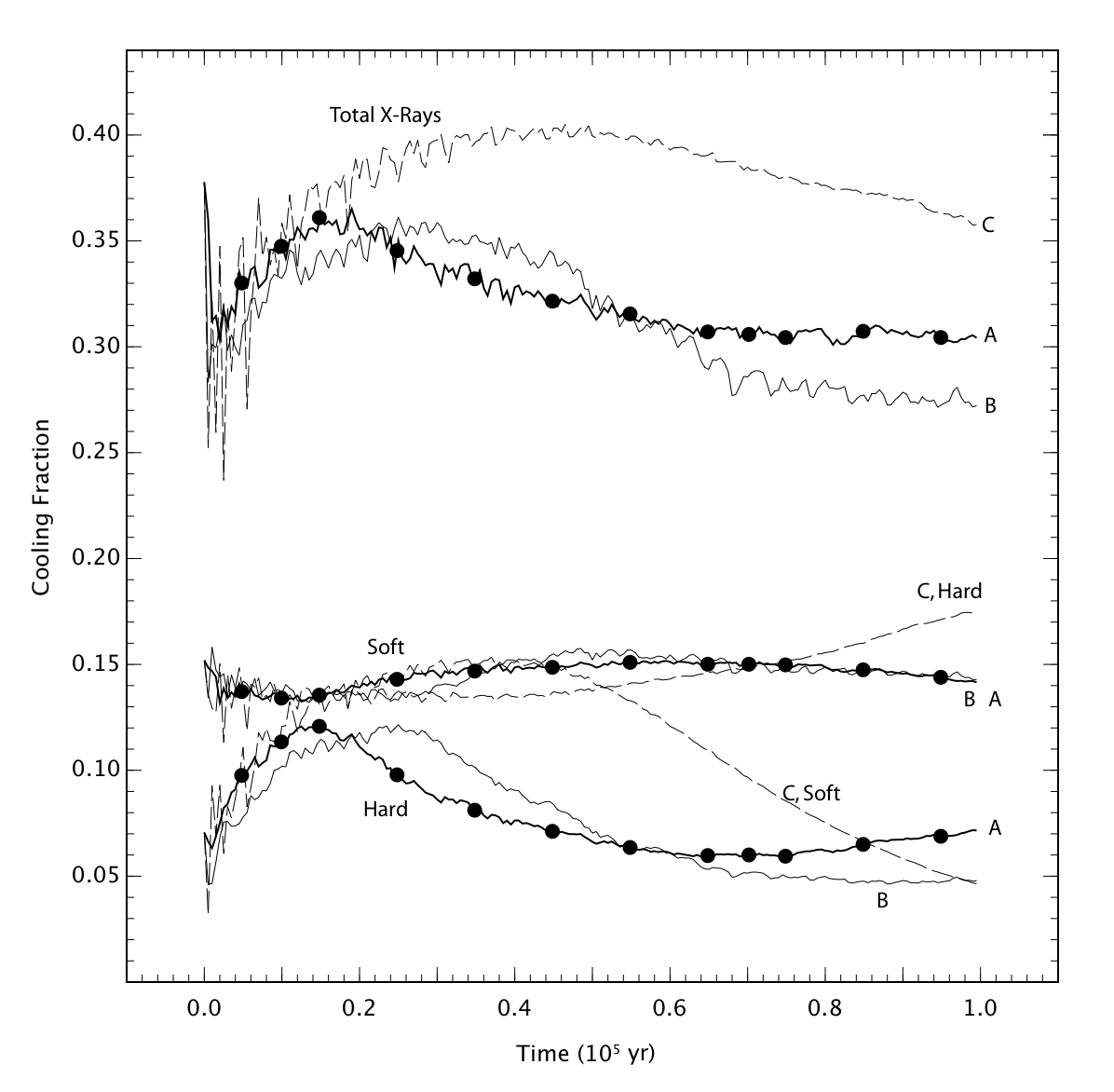}
\caption{\footnotesize The efficiency of the soft ($0.1-0.75$keV) and hard ($1.5-10$ keV) and total  X-ray emission as a fraction of the total thermal cooling of models {\bf A} (heavy line) , {\bf B} (fine line) and {\bf C} (dashed line).  The times of the snapshots used in the detailed description of model {\bf A} are indicated by filled circles.  }
\label{f:L_fX}
\end{center}
\end{figure}

Figure \ref{f:L_fX} shows the corresponding fractions of the total cooling in X-rays and in the soft and hard bands.  In all the models, the X-ray fraction of the total cooling rises to 35--40\%, until the bubble clears the disk. It then declines as the overall cooling becomes increasingly dominated by the very hot main bubble. Nevertheless, the disk continues to contribute a significant percentage of the cooling throughout the simulation.  In the disk dominated soft X-rays, the efficiency in {\bf A} and {\bf B} are remarkably steady at 15\% of all cooling, at all times, where the uniform model {\bf C} shows stronger evolution as the disk is swept away in a strong hot shock.

\clearpage

\section{Mass Redistribution}

The interaction of the jet with the dense warm disk causes a redistribution of mass as a result of the momentum imparted to the warm gas. We examine this process here in detail for two purposes: (1) Too ascertain the potential role of such jet--disk interactions on the feedback of black holes on forming galaxies. and (2) To examine the continued presence of gas in the nuclear regions of radio galaxies well after the passage of the jet to large distances from the core. We examine the redistribution of gas here in all three models {\bf A}, {\bf B} and {\bf C} (see \S\ref{s:simpars}) with the particular emphasis on examining the effect of the greater porosity in Models {\bf A} and {\bf B}, compared to {\bf C}.

\subsection{Morphology}

\begin{figure}
\begin{center}
\includegraphics[width=5in]{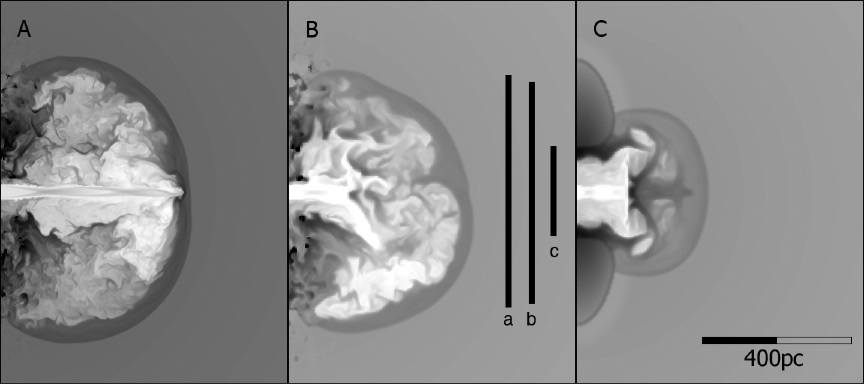}
\caption{{\footnotesize A side by side comparison of the density of models {\bf A} , {\bf B}  and {\bf C}  at $t =75$~kyr. The images are of log density in a slice through the middle of the simulation.  Bars marked a, b and c indicate the extend of the outer blast wave in the central plane of the disk, representing 610, 592 and 240~pc respectively.}}
\label{f:ABC_dens}
\end{center}
\end{figure}

Figure~\ref{f:ABC_dens} is a side by side comparison of the density of the three models at 75~kyr. As expected the dynamical states of the simulations are quite different. Model~{\bf A} is at the end of the jet break-out phase; the jet is about to pierce the bubble that has been inflated during the flood and channel phase. As a result of the higher density of obstructing clouds, Model~{\bf B}  is still in the flood and channel phase and is pushing aside the last remaining cloud in its path. In model~{\bf C} the only way for the jet to escape from the dense environment is for the jet and its associated cocoon to push out a dense plug of gas approximately seven times the diameter of the jet. As can be seen from the image, the morphology is much more symmetric and we are unaware of any real radio galaxy that resembles model {\bf C} .

\subsection{Mass Transport, Jet --  Disk Feedback}
\label{s:transport}

One of the main motivations for examining jet--ISM interactions in detail is to examine the importance of black hole induced feedback processes in galaxy formation. Here we consider the mass that is driven to larger scales as a result of the interaction of the jet with the turbulent disk. In order to facilitate the analysis, we have divided the computational domain into a nested series  of rectangular \emph{zones}, and further define \emph{regions} as the difference between successive zones. Integration of the density in each zone give the zone masses as a function of time, and the differences between successive zone integrals allow us to monitor the transport of material from region to region.  The zones are rectangular for simplicity of integration, although the regions between them may not be.  Figure \ref{f:masszones} shows the regions graphically for a late time density snapshot of Model~{\bf A}.

In describing the locations of the various zones and their sizes, we remind the reader that the $x$-coordinate is in the jet direction and the $y$ and $z$-coordinates are transverse to the jet. The center of the lowest plane of each zone is the origin of the coordinate system. In the follwing descriptions of the zones, the extents are given in the $x$, $y$ and $z$ coordinate directions respectively.
\begin{enumerate}
\item[Zone 1] $64 \times 256 \times 256$~pc.  The first zone, which is also the first region, covers the inner part of the disk around the jet inlet.  It covers a grid area of 256 by 256~pc in the disk plane, centered on the jet inlet,  and extends to 64~pc above the plane.  Region 1 is rectangular.
\item[Zone 2] $64 \times 512 \times 512$~pc. This region is also in the disk plane and extends to the same height above the disk as zone 1, but covers a larger area of the grid,  (512 by 512~pc).  Region 2 is a square annulus between zone 1 and zone 2 and provides a measure of the mass in the outer disk.
\item[Zone 3] $128 \times 768 \times 768$~pc. This zone covers the entire disk, and extends to twice the height of zones 1 and 2. It contains both disk and hot amtosphere.  Region 3, between zone 3 and zone 2, is where material ejected from the disk regions 1 and 2 first appears. Region 3 is not a simple rectangular prism, but forms  a  `cap' above the disk.
\item[Zone 4] $256 \times 1024 \times 1024$~pc. Zone~4 covers the first quarter of the computational volume in the $x$ direction, and the whole $y-z$ plane.  Region 4, between zone 4 and zone 3, comprises the lower part of the hot atmosphere, and does not initially comprise disk material.
\item[Zone 5] $512 \times 1024 \times 1024$~pc. This zone comprises the entire first half of the computational volume. The region between zones 5 and 4 is region 5 -- the mid region of the  hot atmosphere, in the range 0.256 -- 0.512~kpc above the plane.
\item[Zone 6] $1024 \times 1024 \times 1024$~pc. This is the entire grid.  Region 6, zone 6 $-$ zone 5, encompasses the upper half of the hot atmosphere more than 0.512~kpc above the disk plane.
\end{enumerate}

\begin{figure}
\begin{center}
\includegraphics[width=5in]{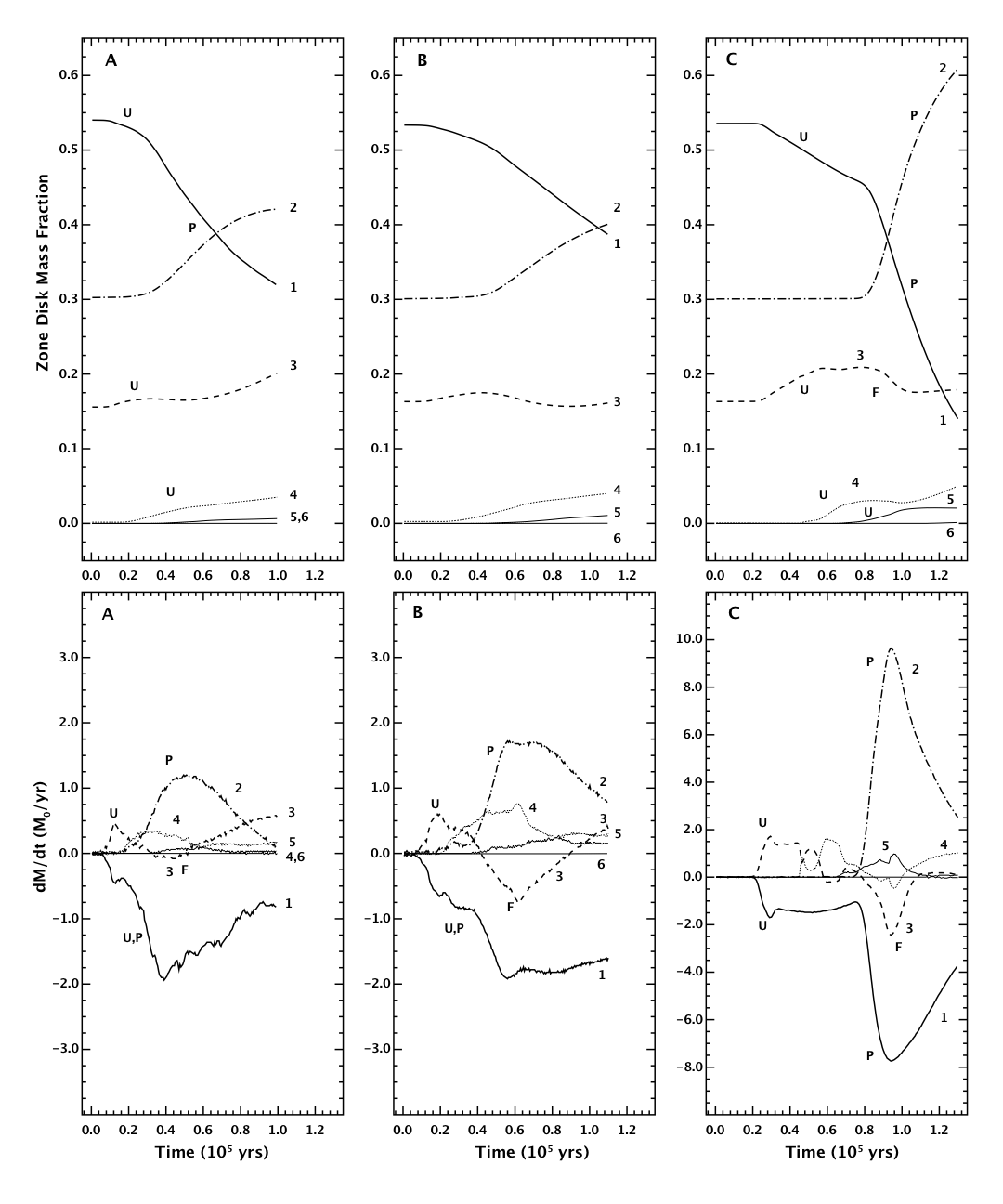}
\caption{{\footnotesize  Mass fractions and derivatives in regions 1-6, for model {\bf A} (left), {\bf B} (center) and {\bf C} (right).  The upper panels show the fraction of disk material, by mass ,in each of the 6 regions, labelled by region number.  The letters {\bf U}, {\bf P} and {\bf F} indicate the dominant direction of mass transfer on the curves. {\bf U} indicate periods of '`uplift',  such as their early transfer from region 1 into region 3 .  {\bf P} indicate in plane movement, such as the strong transfer from region 1 to region 2.  {\bf F} indicates where material is predominately falling back to a lower altitude zone.  The lower panels show the evolution of the mass transfer rates in M$_\odot$ per year, with the same labelling as the upper panels.  Note the lower right panel for model {\bf C} has a larger rate vertical scale then the other two models. }}
\label{f:mma}
\end{center}
\end{figure}

\begin{figure}
\begin{center}
\includegraphics[width=5in]{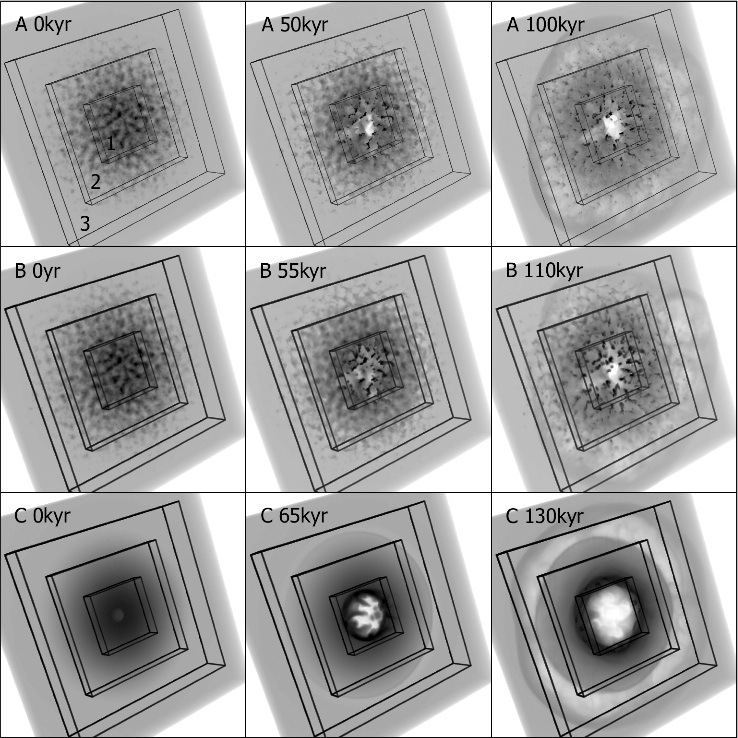}
\end{center}
\caption{{\footnotesize  These images show the distribution of gas density at the beginning, midpoint and final state for the three models in zone 4, {\em i.e.} just the lower 256~pc ($1/4$) of the grid, excluding the upper atmosphere regions 5 and 6 for clarity.  The endpoints were determined by halting the simulation once the jet reached the far edge of the computational grid.  The location of regions 1-3 are shown.}}
\label{f:abc_zones}
\end{figure}

Mass redistribution in the simulations is summarised in  Figures~\ref{f:mma} and \ref{f:abc_zones},  and Table~\ref{t:Zone_mass}. The upper panels in Figure~\ref{f:mma} show, for each model, the disk mass fraction, defined as the fraction of mass in each region occupied by matter originally in the warm disk, as a function of time. The lower panels show the time rate of change of mass in each region as a function of time.   Region 3 shows a range of processes, uplift from zone 1 and fallback to zones 1 and 2, and the mass in region 3 is variable.  The upper atmosphere regions, 5 and 6, see very little of the disk material by the end of the simulations, and even with significant movement of the disk in model {bf C}, 2\% or less ends up above 500~pc from the disk plane.   The mass exchange rates in models {\bf A} and {\bf B} are similar, peaking at around 2 M$_\odot$ per year in the plane of the disk between region 1 and 2, while the monolithic disk shock in the uniform test model {\bf C} generates rates up to 5 times greater.  Models {\bf B} and {\bf C} have same resolution and parameters except for the non-uniformity of {\bf B} suggesting that any quantitative models of mass transfer in jet-host feedback models will need to take non-uniformity into account, and more constraints and detailed knowledge of the non-uniformity in real radio galaxy hosts is needed before truly realistic transport models can be computed.
Table~\ref{t:Zone_mass} summarizes, for each model, the initial total and disk mass and the initial and final disk mass fractions.

\begin{table}
\begin{center}
\begin{tabular}{rlllllll}
\hline
\hline
&\multicolumn{7}{c}{Initial Total Masses ($10^6$ M$_\odot$)}\\
\multicolumn{1}{c}{Model}&\multicolumn{1}{c}{R1}&\multicolumn{1}{c}{R2}&\multicolumn{1}{c}{R3}&\multicolumn{1}{c}{R4}&\multicolumn{1}{c}{R5}&\multicolumn{1}{c}{R6}&\multicolumn{1}{c}{Total}\\
\hline
{\bf A}&0.253&0.145&0.118&0.165&0.208&0.314&1.203\\
{\bf B}&0.506&0.287&0.193&0.166&0.208&0.314&1.673\\
{\bf C}&0.504&0.283&0.184&0.165&0.208&0.314&1.657\\
\hline
&\multicolumn{7}{c}{Initial Disk Mass ($10^6$ M$_\odot$)}\\
\multicolumn{1}{c}{Model}&\multicolumn{1}{c}{R1}&\multicolumn{1}{c}{R2}&\multicolumn{1}{c}{R3}&\multicolumn{1}{c}{R4}&\multicolumn{1}{c}{R5}&\multicolumn{1}{c}{R6}&\multicolumn{1}{c}{Total}\\
\hline
{\bf A}&0.252&0.141&0.073&0.001&0.000&0.000&0.467\\
{\bf B}&0.505&0.285&0.155&0.002&0.000&0.000&0.947\\
{\bf C}&0.504&0.283&0.153&0.001&0.000&0.000&0.941\\
\hline

&\multicolumn{7}{c}{Initial Disk Mass Fractions}\\
\multicolumn{1}{c}{Model}&\multicolumn{1}{c}{R1}&\multicolumn{1}{c}{R2}&\multicolumn{1}{c}{R3}&\multicolumn{1}{c}{R4}&\multicolumn{1}{c}{R5}&\multicolumn{1}{c}{R6}&\\
\hline
{\bf A}&0.540&0.308&0.159&0.002&0.000&0.000&\\
{\bf B}&0.534&0.301&0.163&0.003&0.000&0.000&\\
{\bf C}&0.536&0.301&0.163&0.001&0.000&0.000&\\
\hline
&\multicolumn{7}{c}{Final Disk Mass Fractions}\\
\multicolumn{1}{c}{Model}&\multicolumn{1}{c}{R1}&\multicolumn{1}{c}{R2}&\multicolumn{1}{c}{R3}&\multicolumn{1}{c}{R4}&\multicolumn{1}{c}{R5}&\multicolumn{1}{c}{R6}&\\
\hline
{\bf A}&0.325&0.428&0.205&0.036&0.007&0.000&\\
{\bf B}&0.386&0.401&0.161&0.040&0.011&0.000&\\
{\bf C}&0.140&0.609&0.179&0.050&0.020&0.001&\\
\hline
\end{tabular}
\end{center}
\caption{Model Region Mass Distributions.}
\label{t:Zone_mass}
\end{table}

Figure~\ref{f:abc_zones} shows a projection of the density in the first 3 regions at selected epochs.  It illustrates how the single disk shock in the uniform model, {\bf C}, efectively sweeps
the inner regions of the disk clear of high density gas, in contrast to the non-uniform models {\bf A} and {\bf B}.

Three key points about  the mass redistribution are evident from the figures and table:

\subsubsection{Mass advection efficiency with a radiative critical density}
\label{s:critical _density}

\begin{figure}
\begin{center}
\includegraphics[width=5in]{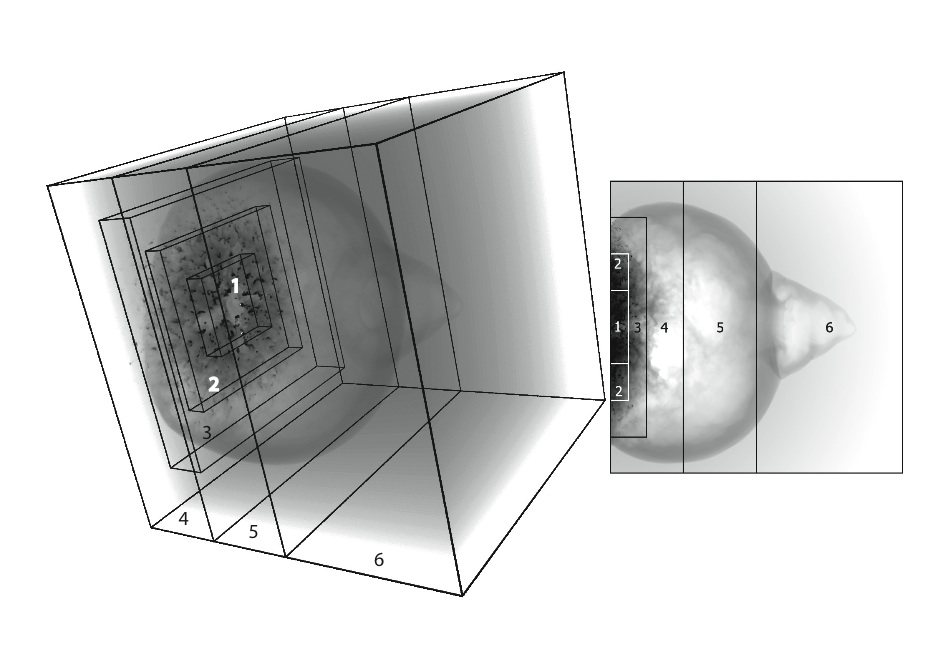}
\end{center}
\caption{{\footnotesize A 3D perspective and side view of the arrangement of the regions in the mass integration analysis.  Each zone encompasses the preceding zones, each being a rectangular integration box, and the labeled regions are defined at the difference between each successive pair of zones, with region 1 and zone 1 being the same.} }
\label{f:masszones}
\end{figure}

From the initial and final mass fraction in table \ref{t:Zone_mass}, overall the two clumpy models ({\bf A} and {\bf B}) are similar, with a significant fraction of disk material remaining in the inner region~1 at the end of the simulation (see also Figure \ref{f:masszones}).  The relative mass remaining in region 1 is somewhat greater in Model {\bf B} compared {\bf A}. This may be consistent with the notion of a critical density,  $\rho_{\rm cr}$, above which the traversing shocks become locally fully radiative. This idea may be explained in the following way: The dynamics of shocked material depends upon the extent to which a shock is fully radiative. In fully radiatively shocked clouds cooling is so important that a dense skin forms around the clouds preventing ablation of gas. This is expected to occur in dense clouds where the cooling time scale is very short since it is inversely proportional to the density. At the other extreme, in low density regions, the heating and subsequent expansion of shocked gas makes ablation of cloud material relatively easy. Thus it is not surprising that there would be a critical density defining the boundary between these two regimes. 

Let us consider the residual amounts of gas in Models A and B. In Model A, approximately 60\% of the gas remains at late times. Given the mean density $\mu = 10 \> \rm cm^{-3}$, an unaffected mass fraction $1-\phi_M(\rho_{\rm cr}/\mu)$ (see equation~\ref{e:lnMassInt}) corresponds to a critical mass density of approximately $15-20 \> \rm cm^{-3}$. In Model~B, approximately 72\% remains and the mean density $\mu \approx 20 \> \rm cm^{-3}$. By the same reasoning this residual mass fraction corresponds to a critical density of approximately $20-22 \> \rm cm^{-3}$. 

 While the detailed spatial distributions, timing, and resolution may be expected to affect the exact fractions, the similarity of the apparent critical density in {\bf A} and {\bf B} suggests that as long as the material above a radiative critical density remains in dense, un-connected, clumps, the local non-gaussian log-normal statistics of the density may influence the behavior of the global system.  The notion of a critical density is developed further in \S~\ref{s:single-point_stats}.
 
\subsubsection{ Inefficient transport of disk ISM to outer hot atmosphere}

In all cases, about 83--85\% of the gas starts in zones 1 and 2 combined, and 75--78\% remain in the two zones at the end of the simulation;
only 4--7\% of the disk gas is transported to the higher regions 4, 5 and 6.  Trace amounts of disk material are transported to region~6 largely in the form of material diffused into the very hot radio plasma. This fraction is highly uncertain since it may be dominated by numerical diffusion rather than physical transport.  Despite the fact that the main blastwave crosses the entire disk, and locally {\em many} shock crossing timescales pass, the majority of the warm clumpy disk ISM is not transported out of the plane of the disk, and the jet outburst is poor at `clearing out' the disk.

\subsubsection{Transport of disk ISM within the disk}

In the uniform model~{\bf C}, 75\% of the disk mass initially in region 1 is swept to region 2, and as the circular empty region increases we expect that essentially all of the inner disk would be swept clear in another $20-40$~kyr.  In models {\bf A} and {\bf B}  only 30--40\% of the disk in region~1 is transported to region~2, and the timescale for clearing out region ~1 is becoming greater than 150,000 years (assuming constancy of the final mass loss rates).  It is therefore likely that dense material remains in region ~1 for a longer time in models {\bf A} and {\bf B} compared to model {\bf C}.  

The clumpy nature of models {\bf A} and {\bf B} significantly extend the lifetime of material in the inner zones of the outbursts, showing that in the cirumnuclear region of an active galactic nucleus, the inhomogeneity of gas in the inner region makes it possible to retain gas there, maintaining a supply of accreting gas to the black hole for longer than models with homogeneous gas distributions may suggest. The relevant timescale for the duration of the outburst is determined primarily by  the accretion timescale. 

The maintenance of gas in a disk-like distribution in the central regions is also relevant to observations of radiating gas in radio galaxies that has clearly survived beyond the passage of the jet through the disk. In this context we are mainly thinking of the X-ray observations of the inner regions of  Cygnus~A where the source may be aged  $10^7$ years or more in the current outburst, and yet retains a dense and highly structures warm atomic and molecular ISM in it's inner kiloparsec regions.
\citep{wilson00a,wilson06a}.

\section{Evolution of the density structure in the interstellar medium}
\label{s:density_evolution}

\subsection{Evolution of the single-point statistics}
\label{s:single-point_stats}

As described in \S~\ref{s:warm_ISM} the simulations in Models~{\bf A} and 
{\bf B} are initialized with an ISM structure whose one-point statistics are described by a log-normal distribution and whose two-point structure is described by a power-law in Fourier space. It is interesting to see how this structure evolves in time. Indeed, the time evolution underlines some of the points that we have made above concerning the evolution of high and low density regions and the existence of a critical density.

\begin{figure}
\begin{center}
\includegraphics[width=5in]{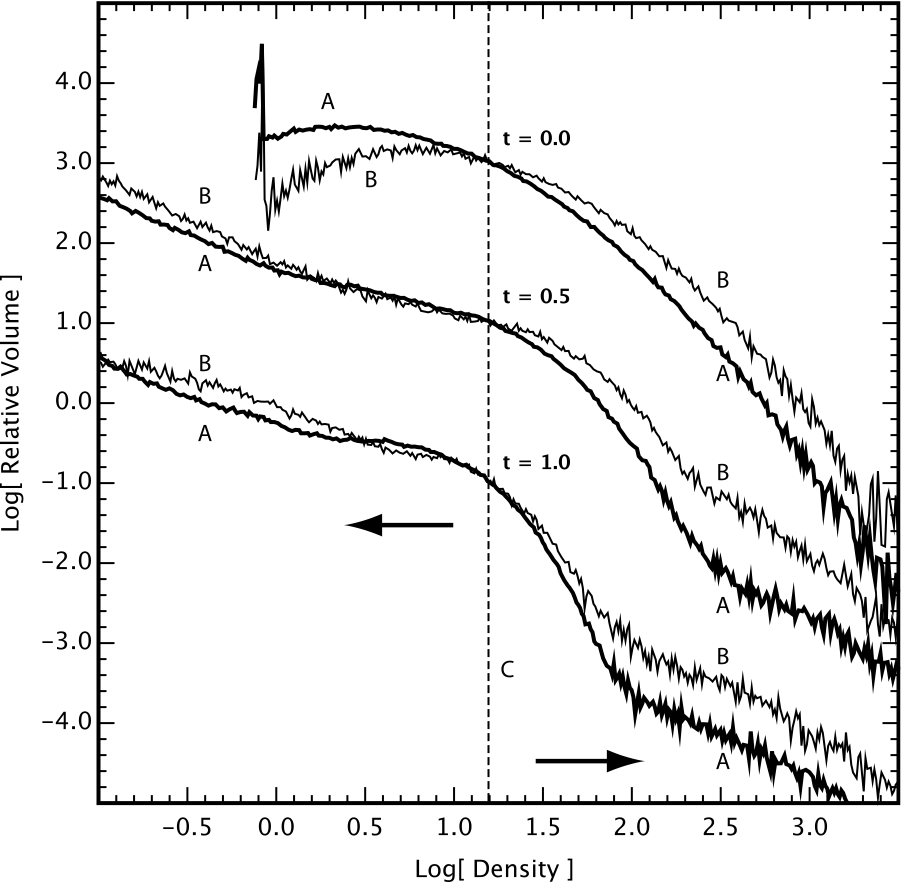}
\end{center}
\caption{{\footnotesize Histograms of the distribution of density in models A and B within the innermost $ 64 \times 256 \times 256$ parsecs region occupied by the disk. These histograms are presented at three different parametric times, $t = 0,\> 0.5$ and 1. The parametric time $t=1$ corresponds to the times near the end of the simulations where similar dynamical states are attained, i.e. at times of 100~kyr and 110~kyr for models {\bf A} and {\bf B} respectively. The parametric time $t=0.5$ represents the halfway point in each simulation. }}
\label{f:density_evolution}
\end{figure}

Figure~\ref{f:density_evolution} shows the evolution of the density distributions within the innermost $64 \times 256 \times 256$ parsecs region occupied by the disk. The histograms represent the volume of gas in a given density range and are presented at three different dynamical times. The histograms are scaled vertically to agree at $\log n = 1.2$ (the dashed line labeled `C' in Figure~\ref{f:density_evolution}) since, as becomes clear in the discussion below, this defines an approximate pivot point in the evolution of the density. For clarity each successive histogram is displaced vertically by 2 dex. Hence, while the plotted histograms do not represent the absolute values of the volume filled by gas of a specific density they represent \emph{relative} volume at each density and this is all that is required in the following.

With reference to Figure~\ref{f:density_evolution} we identify the following principal features of the evolution of the desnity distribution:
\begin{enumerate}
\item At $t=0$ the histograms of Models {\bf A} and {\bf B} show the expected shape of truncated log-normal distributions with some low density spikes at the extreme left of each distribution corresponding to the inlet of jet gas at the origin and the pressure cut-off limit, which replaces low pressure ISM regions with hot halo gas. 

For $\log n < 1.2$ (i.e. to the left of C) Model~{\bf A} has relatively more low density gas and Model~{\bf B} has relatively more high density.

\item At $t=0.5$ (corresponding to times of 50~kyr and 55~kyr in Models~{\bf A} and {\bf B} respectively) the density distributions for $1.2 \la \log n \la 2.3$ in each model are similar to the initial distributions. We attribute this to minimal processing of most of the volume at this stage of the simulations.  Nevertheless there are some notable new features. High density tails have developed, which are the result of the dense cores formed by radiative shocks. At the same time the distributions are enhanced at low density as a result of non-radiative low density gas which has been shocked and advected by the the disrupted jet and the associated energy-driven bubble. 

The distributions to the left of C show that the models {\bf A} and {\bf B} become very similar, while the curves to the right of  C remain similar to the initial conditions, albeit with the appearance of the high density (low volume) tail.  This was the basis for deciding to normalize the curves at C for comparison.  The density in the vicinity of C may represent a critical density related to cooling that affects advection.  However as the curves between  C and approximately 2.3 have not changed, the location of C may represent a shock crossing time with the densest initial core not yet being fully shocked at this stage.

\item At $t = 1.0$, The distributions to the left of C remain approximately the same as at $t = 0.5$, indicating that a quasi--steady state  has been set up as far as the advection of the ISM is concerned.  The distributions for Model~{\bf A} at 100~kyr and Model~{\bf B} at 110~kyr are similar to the left of C, except that model {\bf A} shows some sign of reestablishing a log-normal like distribution for $0.5 \la \log n \la 2.0$, which is narrower and lower than the initial distribution.  Whether this is due to a turbulent cascade forming in the low density outflow is unclear.

The curves to the right of C are also becoming more similar, indicating that more complete processing of all the high density has taken place, and that in the range $1.2 \la \log n \la 2.0$ the density distribution is tending towards a single distribution.  Since the distributions pivot at approximately $\log n = 1.2$ we tentatively identify $n \approx 16 \> \rm cm^{-3}$ as the critical density, initially discussed in \S~\ref{s:critical _density} related to the radiative shock velocities in the fractal ISM.  Above the critical density the very high density shocked gas cools rapidly becomes denser and is retained; below $\log n = 1.2$ the low density gas does not cool quickly and is swept away.
\end{enumerate}

In conclusion, Figure~\ref{f:density_evolution} is consistent with there being a critical density of about 16 particles per $cm^3$, above which the advection of shocked gas is inefficient. Since model {\bf B} has a higher mean density, this threshold bounds a larger fraction ($z(\rho)$ of the underlying log--normal distribution, resulting in a larger residual fraction of gas at $t = 1.0$.

\subsection{Evolution of the Fourier spectrum}

\begin{figure}
\begin{center}
\includegraphics[width=5in]{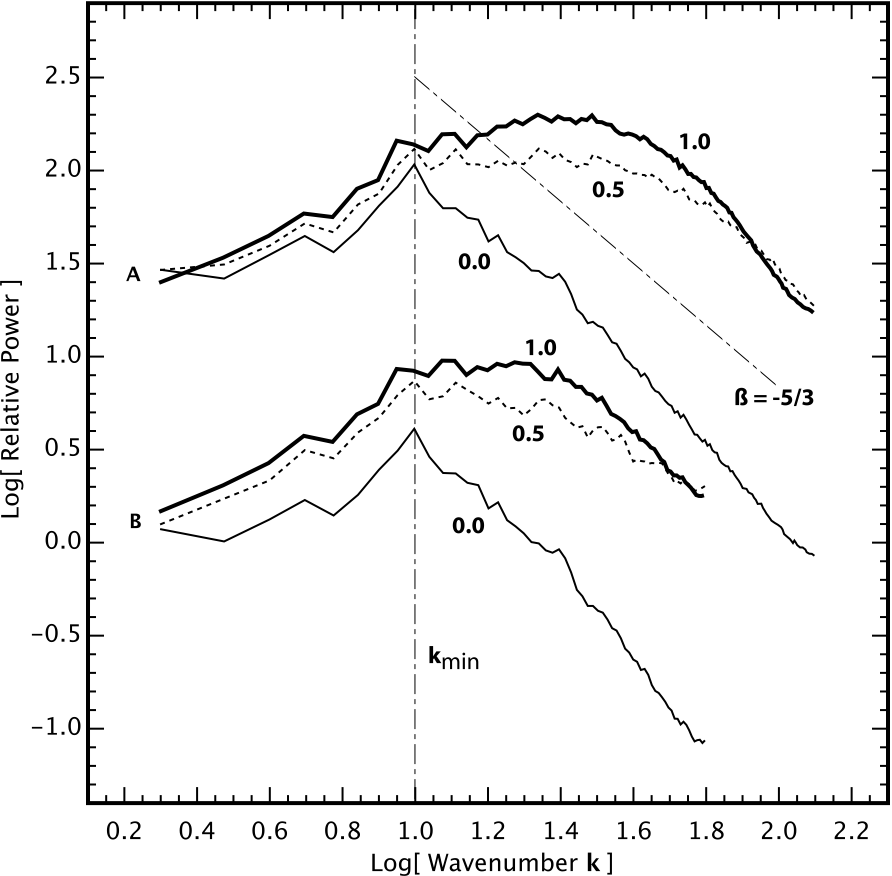}
\end{center}
\caption{{\footnotesize  The evolution of the density in Fourier space. The curves represent the averaged power spectrum of the density at parametric times  $t = 0,\> 0.5$ and 1 respectively. }}
\label{f:fourier_evolution}
\end{figure}

Figure~\ref{f:fourier_evolution} shows the spectral power density of the mass density in the same region as above \emph{viz.} the innermost $ 64 \times 256 \times 256$ parsecs region occupied by the disk for both Models~{\bf A} and {\bf B}. The same parametric times are also represented. The horizontal wave number scale is defined in terms of $k/k_{\rm min }$ where $k$ is the magnitude of the wave number and $k_{\rm min}$ is the minimum wave number consistent with the size of the computational domain. The larger extent in $k$-space of the power spectrum in Model~A is the result of the factor of two greater resolution.

The Model~{\bf A} and Model~{\bf B} power spectra are translated vertically by 2 dex for clarity. However, there is no other normalization of the spectra.  

At $t=0$ both models of course show the expected high wave number $\beta = -5/3$ slope that is imposed on the initial fractal data cube. In discussing the evolution of the spectra, it is useful to discuss Model~{\bf B} first. The most obvious feature in the power spectrum is a growth of intermediate scale structure for $1.0 \la k \la 1.4$ corresponding to regions on the size of 5 to 13 cells. We attribute this to the propagation of shocks into the dense gas causing filamentation and ablation.

For $k \ga 1.4$ the $\beta = -5/3$ slope is maintained possibly indicating the establishment of the inertial region of a turbulent cascade. However, we cannot be definitive about this without further investigation of the spectral properties of the numerical solution, which would take us beyond the scope of this paper.

Model~{\bf A} shows very similar characteristics. However, the new intermediate structure extends further in $k$-space, over a range $1.0 \la k \la 1.9$ corresponding to regions of the size of approximately 3 to 26 cells. Again we attribute this feature in the power spectrum to shock filamentation and ablation. In this case however, the lower mean density of the warm gas allows these processes to propagate to smaller scales. In this case there is the merest hint of a power-law spectrum at larger wave numbers since the broad feature tat we have been discussing almost to the Nyquist limit.

\section{The compact symmetric object 4C31.04}

In this section we consider the application of the insights that we have gained form these simulations to a specific Compact Symmetric Object -- 4C31.04.

Many powerful Fanaroff-Riley Class~2 (FR2) radio galaxies fit the standard paradigm of two lobes situated on either side of an active nucleus with obvious hot spots, which are assumed to be the termini of relativistic jets \citep{scheuer74a, blandford74a}. Moreover, simulations of supersonic jets propagating through an homogeneous interstellar medium seem to confirm this simple picture. However, on closer examination, many sources are \emph{not} consistent with it. In particular, examples can be  drawn from the classes of compact radio sources such as Gigahertz Peak Spectrum (GPS) sources, Compact Steep Spectrum (CSS) sources and Compact Symmetric Objects (CSOs) which do not fit the standard model of jet -- compact hotspot -- backflowing cocoon. Our proposition is that departures from the standard picture reveal additional physics in the evolution of radio sources that has been neglected to a large extent and that the simulations presented here fill in at least some of the gaps in our understanding of these sources.

The compact symmetric object 4C31.01 \citep{cotton95a} is a case in point. Referring to the most recent VLBA and Merlin images together with the spectral index map in \citet{giroletti03a}, one can see that the plasma in the western lobe does not appear to flow back from the hot spot in that lobe. There is also an extended region of flat spectral index extending to the North away from the hot spot. On the eastern side the hot spot is well recessed and there is a region of flat spectral index extending to the south.

As far as the distribution of thermal gas is concerned, \citet{conway96a} has inferred the existence of a disk in 4C31.01 viewed in absorption against the radio source continuum. The opacity image is more extended on the eastern side compared to the western and Conway interprets this as an aspect dependence of the line of sight through the disk. However, the extent of the region of non-zero opacity could also be the result of an asymmetric distribution of gas in the core of 4C31.04.

Hence in 4C31.04, the physical environment is similar to that established for this simulation, save for the fact that the physical scale of the simulation is approximately a factor of 10 larger. Nevertheless, we suggest that the qualitative physical interactions that we have identified in the simulation define the important features of jet-disk interactions and that the precise scale is secondary. One needs to make the qualification that the density of the warm medium has to be such that the evolution of that medium is similar, such as having similar volume filling factors of the gas where the density is greater than the critical radiative density $\rho > \rho_{\rm cr}$.  Here in models {\bf A} and {\bf B} the radiative gas clouds appear to consist of gas denser than the mean density, and would have a volume filling factor of $1-\phi_V(\mu) \sim 0.25$ or less.

If the critical density is lower down the density distribution,  that is clouds are too dense everywhere,  they could form an impenetrable obstacle to the flow that are difficult to move;  if the gas with high density is a very small fraction of the medium,  the bulk of the medium will be easily blown away,  and not obstruct the radio plasma much at all.   Hence, it would be interesting to tailor the physical scale and density distribution of a simulation to 4C31.04. However, it is worth noting the following strong similarities between model~{\bf A} and the morphology of 4C31.04, which we believe shed some light on the physics of that source.  We also note the within the restricted scaling allowable for model {\bf A}, the  disk/warm ISM mass scales quadratically with $x_0$, and time scales linearly.  
\begin{enumerate}
\item On the western side, the outer edge of the radio lobe is straight and perpendicular to the direction of the radio axis. The radio morphology is similar to the 70 and 75~kyr (7--7.5~kyr, for $x_0 = 0.1$~kpc) synthetic radio images in Figure~\ref{f:inu_B}.
\item  On the Eastern side the structure of the lobe and the recessed hotspot suggest comparison with the 35 -- 65~kyr~epochs of model~{\bf A}. (3.5--6.5~kyr).
\end{enumerate} 

Our speculation is that the Western lobe of 4C31.04 is near the end of the jet breakout phase after an interaction with a dense warm ISM with a mass $< 10^4$~M$_\odot$, with the hot spot about to pierce the radio bubble and commence the formation of a classical double source. The eastern lobe, on the other hand appears to be in a slightly earlier dynamical phase having formed a bubble within which the jet terminus is propagating to the edge. The different dynamical phases of the lobes on both sides of the source may be related to an asymmetric disk-like distribution of dense gas on both sides of the source -- consistent with the \citet{conway96a} HI opacity images.

\begin{figure}
\begin{center}
\includegraphics[width=5in]{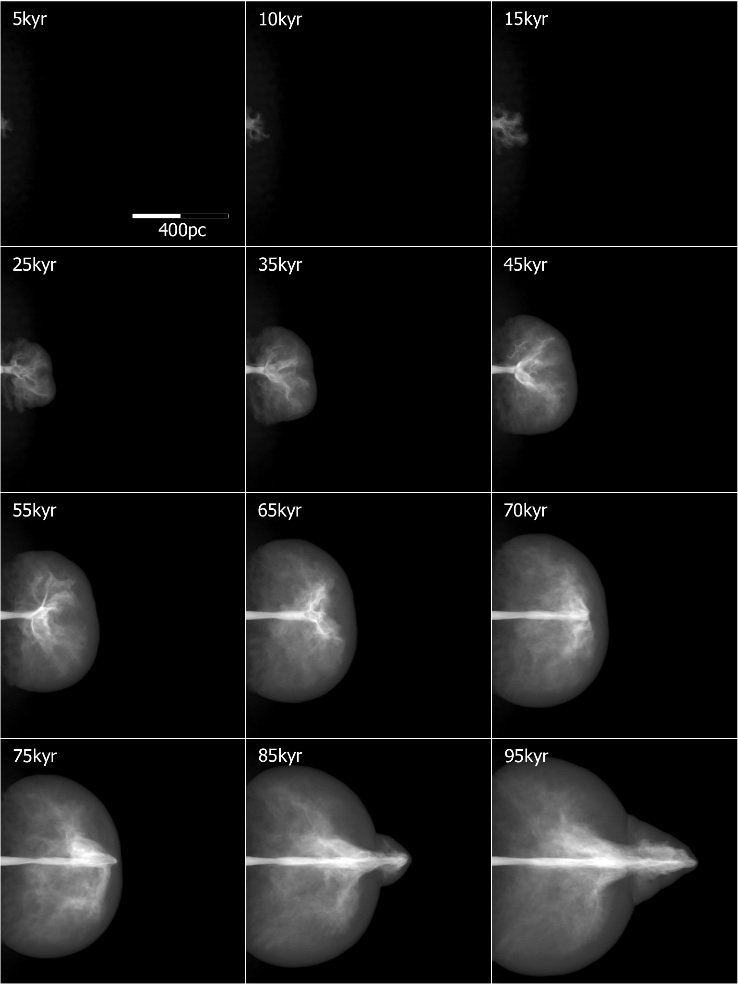}
\caption{\footnotesize The linear RMS line-of-sight velocity magnitude, normalised to the maximum jet velocity at late times, white is fast, black slow.}
\label{f:velm}
\end{center}
\end{figure}

Another feature of 4C31.04 which is consistent with a scaled model {\bf A} is the region of flat spectral index in the Western lobe following the bright ridge line to the North.
The series of snapshots of the velocity magnitude in Figure~\ref{f:velm} emphasizes the filamentary nature of the jet velocity. Moreover, near the time when the jet is about to pierce the bubble, there are filaments of significant velocity magnitude perpendicular to the jet. Dissipation of this velocity in shock induced particle acceleration could lead to flat spectral index features similar to those observed in 4C31.04.

\citet{giroletti03a} highlighted a paradox with the observations of proper motions in 4C31.04. Four of the brighter components (including the Western hotspot) are observed to have mildly relativistic velocities indicating a kinematic age for the source $\approx 550 \> \rm yr$. However the spectral age of the source (assuming equipartition conditions) $\approx 3-5,000 \> \rm yrs$. \citet{giroletti03a} discuss a number of possible reasons for this discrepancy including wandering of the jet direction and inappropriateness of the equipartition assumption. However, we suggest on the basis of model~{\bf A} that the source has been in the spherical bubble phase for about 3-5,000~yr, that persistent collimated jets are a relatively recent phenomenon ( $< 1$~kyr )and that the high hot spot velocities represent the phase where jet breakout is about to occur. This differs from the ``wandering jet'' model proposed by \citet{giroletti03a}. Their model implies that recognisable jets have been active in the source for some time; our model implies that the jets have been frustrated for a large proportion of the source lifetime.

In support of this model we estimate the jet energy flux and ambient density that is implied by the assumption that the lobes of 4C31.04 are jet driven bubbles. The theoretical basis for these calculations is the equations presented in \citet{bicknell96a} for the radius and pressure of a bubble filled with relativistic plasma deposited by a jet. One can easily invert these equations to determine estimates for the jet energy flux, $F_{\rm E}$, and ambient density, $\rho_{\rm a}$, in terms of the bubble radius, $R_{\rm b}$ and energy density $\epsilon_{\rm b}$. These equations apply to a bubble in which the magnetic field is neglected compared to the plasma energy density and we use minimum energy estimates below as fiducial estimates for the bubble energy. However, the discrepancy is not serious since a disordered magnetic field behaves like a gas with a polytropic index of $4/3$ and the minimum in the energy density is a shallow one when the independent variable is taken to be the ratio of magnetic to plasma energy density.

We have for the jet energy flux and ambient density:
\begin{eqnarray}
F_{\rm E} &=& \frac {32 \pi }{25} \, 
\frac{\epsilon_{\rm b} R_{\rm b}^3}{t_{\rm b}} \, ,\nonumber \\
\rho_{\rm a} &=& \frac {25}{36} \, 
\frac {\epsilon_{\rm b} t_{\rm b}^2}{R_{\rm b}^2}  \, ,
\label{e:bubble}
\end{eqnarray}
where $t_{\rm b}$ is the age of the bubble.

\begin{table}
\begin{center}
\begin{tabular}{c c c c}
\hline
Parameter & Unit & West Lobe & East Lobe \\
\hline
\dag$z$          &       & \multicolumn{2}{c}{0.059} \\
\dag$H_0$    & $\rm km \>s^{-1} \> Mpc^{-1}$ & \multicolumn{2}{c}{70} \\
$D_L$    & Mpc       & \multicolumn{2}{c}{253} \\
\dag$\Phi_x$ & mas & 48 & 60 \\
\dag$\Phi_y$ & mas & 88 & 45 \\
\dag$\Phi_z$ & mas & 88 & 45 \\
$V_{\rm L}$  & pc$^3$ & $3.6 \times 10^5$ & $1.5 \times 10^5$ \\
\dag$\nu$    & Hz     & \multicolumn{2}{c}{$1.7 \times 10^9$}  \\
\dag$F_\nu$  & Jy     & 1.366 & 0.961 \\
\dag$a$      &        & \multicolumn{2}{c}{2.2} \\
\dag$\gamma_1$ &      & \multicolumn{2}{c}{10} \\
\dag$\gamma_2$ &      & \multicolumn{2}{c}{$10^5$} \\
$\epsilon_{\rm p}$ & $\rm ergs \> cm^{-3}$ & $1.1 \times 10^{-7}$ &
 $2.4 \times 10^{-7}$\\
$B$        & Gauss & $1.5 \times 10^{-3}$ & $2.2 \times 10^{-3}$ \\
$\epsilon_{\rm tot}$ & $2.1 \times 10^{-7}$ & $4.4 \times 10^{-7}$ \\ 
$R_{\rm b}$ & pc & 44.0 & 33.2 \\
$F_E$ & $\rm ergs \> s^{-1}$ & $4.4 \times 10^{43}$ & $1.5 \times 10^{43}$ \\
$n_{\rm a}$ & $\rm cm^{-3}$ & 0.12 & 0.45 \\
\hline
 \multicolumn{4}{l}{ {\footnotesize Input parameters are denoted by \dag.}}
\end{tabular}
\end{center}
\caption{Parameters for the Western and Eastern lobes of 4C31.04} 
\label{t:bubble}
\end{table}

In order to estimate the minimum energy density and magnetic fields in the eastern and western lobes of 4C31.04 we approximate the lobes as ellipsoids and estimate longitudinal and lateral angular extents, $\Phi_x$ and $\Phi_y$ of the major axes from \cite{cotton95a}. The extent of a lobe along the line of sight, $\Phi_z$, is taken to be the same as the lateral extent of each lobe. The plasma energy density and magnetic field are assumed to be constant. Let 
$F_{\nu,\rm L}$ and $V_{\rm L}$ be the flux density and volume of each lobe, $c_E$ the ratio of particle energy density to electron energy density, 
$\gamma_1$ and $\gamma_2$ the upper and lower cutoff Lorentz factors and $a$ the electron spectral index. The minimum energy magnetic field and corresponding plasma energy density are calculated using the formulae given in \citet{bicknell03c} and the results are summarised in Table~\ref{t:bubble}.

Treating the minimum energy density as a fiducial but indicative value, we use the bubble equations (\ref{e:bubble}) to estimate values of the energy flux and ambient density for each lobe. The radius used is the geometric mean of the three radii estimated from the angular extents. The resulting values are also summarized in Table~\ref{t:bubble}.

There are two features of the parameter estimates that are of immediate interest. The first is the jet energy flux, which in both lobes is approximately $\hbox{a few} \times 10^{43} \> \rm ergs \> s^{-1}$. This is consistent with the overall power of the source $\approx 1.9 \times 10^{25} \> \rm W \> Hz^{-1}$, which is borderline FR1/FR2 (see \citet{bicknell95a}).

The other feature of interest is the ambient number density $\approx 0.1 - 0.6
\> \rm cm^{-3}$. This is typical of the central number densities in the hot atmospheres of elliptical galaxies.  Hence the notion that the lobes in 4C31.04 are just nearing the end of a jet-driven bubble phase is consistent with our ideas on radio emitting elliptical galaxies.

\section{Summary}

This set of simulations is the first in a series of simulations designed to understand in detail the interaction of powerful jets with the inhomogeneous interstellar medium of radio galaxies. Such a medium may occur as a consequence of a cooling flow into the galaxy core, a merger with a gas-rich companion or a halo of clouds associated with early or intermediate stages of galaxy formation. In the simulations we have presented here, we have assumed that the warm inhomogeneous medium has settled into a disk whose structure is described in terms of a model in which the velocity is almost Keplerian and the disk is vertically supported by supersonic turbulent velocity. Supersonic turbulent disks apparently exist in view of the observations of, for example, M87 and NGC~7052 \citep{dopita97a,vandermarel98a}. The investigation reported here represents one of the numerous possibilities that one can consider for the initial configuration. Nevertheless, the investigation of this specific case is important in its own right. 

The main simulation that we have presented, of a jet interacting with a turbulently supported disk, exhibits four interesting phases: 
\begin{enumerate}
\item  A `flood and channel'  dense ISM interaction phase in which a high pressure bubble  leaks through channels in an inhomogeneous medium. 
\item A nearly adiabatic energy driven bubble phase in which the jet, still disrupted by the clouds in the vicinity of the nucleus drives a pseudo-spherical bubble into the galactic atmosphere. 
\item  A jet breakout phase in which the jet collimates and breaks free of the energy bubble. 
\item A classical phase in which the jet starts to establish the characteristic morphology of an FR2 radio source.
\end{enumerate}

Interestingly, the late-time morphology of the radio source still manifests a clear signature of the two earlier phases. The X-ray emission also presents some further insights. Not only do we see X-ray emission from the disk associated with radiative shocks in the early stages and X-ray emission associated with the bow-shock of the radio lobes but we also see persistent X-ray emission from the disk well after the radio bubble has passed through the disk. This is associated with the continued driving of radiative shocks into the disk by the high pressured, synchrotron-emitting bubble.
 
Nevertheless, the peak in the X-ray luminosity occurs during the jet breakout phase, as more of the jet energy flux is directed into the growing, hot adiabatic bubble volume and there is less power available for processing within the dense disk.  The behaviour of the disk is suggestive of a critical cloud density where the mass available for rapid mass advection may be influenced by the log--normal density field in the disk and the range of fully radiative to adiabatic shocks. 

The global movement of disk material throughout the simulation is clearly very different for a uniform disk model compared to non-uniform models and suggests that non-uniformity in a turbulent host galaxy ISM is a {\em key} element in any future quantitative jet feedback, or indeed any galactic feedback, models. 
Another key concept which we have explored to some extent in \SS~\ref{s:critical _density} and \ref{s:single-point_stats} is that of a critical density which governs the fraction of disk material that is advected to large radii. Gas below the critical density is shocked, heated and ablated; gas above the critical density cools quickly and remains in place.
 Deeper knowledge of the ISM structure, turbulence theory and more consistent algorithm physics (MHD, relativistic flows for jet model) and higher resolution will be useful in advancing this field.

The test  simulations that we have presented has been ``fine-tuned'' to some extent in order to exhibit the above characteristics of the interaction of a jet with an inhomogeneous interstellar medium. Jets of higher power quickly break free of the disk of material surrounding the core and do not exhibit an extended flood and channel or jet breakout phases. On the other hand, we are mainly concerned with the interaction between powerful FR2 class jets and their environment. Hence we have opted for a jet power only slightly in excess of $10^{43} \> \rm ergs \> s^{-1}$. Nevertheless, given this restricted parameter range, the simulation exhibits many of the characteristics that one expects of a jet interacting with a disk-like distribution of inhomogeneous clouds. However, in order to manifest the morphological characteristics that we have discussed in this paper, more powerful jets require disks that are more extended in the vertical direction and which probably have a different physical character than the statistically steady equilibrium disks that we have utilized here.

We have suggested that the radio source 4C31.04, classified as a Compact Symmetric Object by \citet{cotton95a} is an example of the evolution that we have studied in detail with these simulations. 4C31.04 is approximately an order of magnitude smaller in size than the strict application of our models would allow. However, the physics of jet-disk interaction, jet-driven bubbles and jet breakout are still relevant. The notion that the lobes represent the final stages of jet-driven bubbles provides a plausible explanation for the morphology of the lobes, especially that of the Eastern lobe with its recessed hot spot. Such a model also provides credible estimates for the jet energy flux and ambient density. It also explains the discrepancy between the ages based upon component proper motions and spectral steepening. The filamentation of the jet during the jet breakout phase also qualitatively explains the regions of flat spectral index in the Western and Eastern lobes.
 
Finally we also remark on the relevance of the phases of radio galaxy evolution that we have identified to Cygnus~A. The radio structure of Cygnus~A is 60 times larger than our model and for the models discussed here to be relevant the equilibrium turbulent disk would need to be scaled similarly requiring unacceptably large turbulent velocities. However, it is not out of the question that a non-equilibrium disk-like configuration could occur in Cygnus A as a result of a merger for example. Hence the structure of Cygnus~A on a scale of 60~kpc may well be understood in terms of similar jet--ISM interactions and similar dynamical phases to those we have studied in detail here.  Recent Chandra X-ray images \citep{wilson00a,wilson06a} show X-ray emission from the expanding bow-shock similar in morphology to the classical phase of Model~A. There are also brighter filaments closer to the centre of the galaxy, the latter associated with emission line gas on a spatial scale of 2-3 kpc \citep{jackson98a, tadhunter03a}. Using the models in this paper as a guide we would interpret the inner filamentary emission as shocked and photionized cooler gas that was previously distributed in a disk-like distribution of approximately 2--3~kpc vertical extent, in the core of the galaxy. The 327~MHz radio image of Cygnus~A \citep{lazio06a} also shows a plume-like extension of emission to the South. Such extensions have often been attributed to the deflection of the backflow from the lobe hotspots. However, this could also be an indication of an earlier jet-driven bubble phase.

\subsection*{Acknowledgments}
We thank the ANU supercomputing facility time assignment committee for allocations of computer time. This research was also funded by ARC discovery project grants DP0345983 and DPDP0664434.



\appendix

\section{The log-normal distribution}
\label{a:lognormal}

In this appendix we summarise properties of the log--normal distribution some of which are used in the main text; others may be useful in related and future work.

We use a log--normal distribution to describe the single point statistics of the density field of our nonuniform ISM.  The log--normal distribution is a skewed continuous probability distribution, for an independent variable $x \geqslant 0$.   Unlike the normal distribution, it has a non-zero skewness, variable kurtosis, and in general the mode, median and mean are unequal.   The log-normal distribution appears to be a nearly universal property of isothermal turbulent media in experimental, numerical and analytical studies ( {\em e.g.} \citet{nordlund99a} , see also \citet{warhaft00}, and \citet{pumir94} ).  A key property ( see for example \citep{mollo73} ) appears to be that the distribution should display intermittency  -- described by high order moments -- a key element in all current turbulence theories.  The statistic most often used is the {\em flatness}, F, ( {\em c.f}\/ \citet{cerutti98}).  This is the 4th moment divided by the 2nd moment squared:
\begin{equation}
F = \frac{\langle [ f - \langle f \rangle ]^4 \rangle}{\langle [ f - \langle f \rangle ]^2 \rangle ^2} \, ,
\end{equation}
and is also equal to the standard {\em kurtosis} statistic plus 3.
For a normal Gaussian distribution $F = 3$, and when $F \gg 3$ the field displays strong intermittency, with sharp increments arising from the long `tail' of the distribution.  With the log--normal distribution $F = \exp[4s^2] + 2\exp[3s^2] + 3\exp[2s^2]$, where $s^2$ is the log--normal variance,  and can be $\gg 3$ for even moderate values of variance.  It is encouraging  that the log--normal distribution is the limiting distribution for the product of multiplicative random increments, in the same way that the normal distribution plays that role for additive random increments. It is thus compatible at least conceptually with a generic cascading process consisting of repeated folding and stretching.

 With a log--normal distribution, the natural logarithm of the ISM density field is a Gaussian which has a mean $m$ and variance $s^2$, the density field itself has a mean $\mu$, and variance $\sigma^2$.  

The cumulative log--normal probability distribution is:
\begin{equation}
D(x) = \frac{1}{2} \left \lbrack 1 + \mbox{\rm erf}  \left(\frac{\ln x - m}{s \sqrt{2}}\right ) \right \rbrack \, ,
\end{equation}
The corresponding probability density function is then,
\begin{equation}
P(\rho) = \frac{1}{s \sqrt{2 \pi} \, \rho} \exp \left \lbrack \frac{-(\ln \rho - m)^2}{2s^2}\right \rbrack \, .
\end{equation}

\section{ Log-normal Distribution and Integral Tables }
\label{a:log-normal-integrals}

The log--normal distribution has a number of properties; we note here the common low order moments of the density distribution in terms of the log--normal parameters $m$ and $s^2$:
\begin{eqnarray}
\mbox{\rm mean}, \mu & = & \exp[ m + s^2/2]  \\
\mbox{\rm variance}, \sigma^2 & = & \mu^2 \, (\exp[ s^2]-1) \\
\mbox{\rm mode}   & = & \exp[ m - s^2] \\
\mbox{\rm median} & = & \exp[ m] 
\label{e:rels}
\end{eqnarray}
with the inverse relations:
\begin{eqnarray}
m &=&   \ln [\mu^2] - 1/2 \,\ln [ \sigma^2 + \mu^2 ] \, . \\
s^2 &=&  \ln[ \sigma^2 + \mu^2] - \ln[ \mu^2] \, .
\label{e:lnrels}
\end{eqnarray}
For our standard values of $\mu = 1.0$ and $\sigma^2 = 5.0$, table \ref{t:summary} gives the numerical values for these modes.

\begin{table}
\begin{center}
\begin{tabular}{ r c c}
\hline
\hline
\multicolumn{3}{c}{$\mu = 1, \, \sigma^2 = 5$}\\
Parameter       & $F(r)$ & $f(r)$\\
\hline
mean & $\mu = 1.0$ & $m \approx -0.895880$\\
variance & $\sigma^2 = 5.0$ & $s^2 \approx 1.79176$\\
median & $1/\sqrt{6} \approx 0.408248$&\\
mode & $\approx 0.0680414$&\\
\hline
\end{tabular}
\end{center}
\caption{Standard Log-Normal Distribution Parameters.}
\label{t:summary}
\end{table}

For a density variable $\rho = F(\vec{r})$, let $\zeta = \phi_V(z)$ be the fractional volume with density less than a threshold, $\rho<z$:
\begin{eqnarray}
\phi_V(z) & = & \int_{0}^{z} P(x) dx \nonumber \\
                 & = & D(z)\nonumber \\
                 & = &  \frac{1}{2} \left \lbrack 1 + \mbox{\rm erf}  \left(\frac{\ln z - m}{s \sqrt{2}}\right ) \right \rbrack  \, .
\label{e:lnVolInt}
\end{eqnarray}
Also, let  $\zeta =\phi_M(z)$ be  the fractional mass with density less than a threshold $\rho<z$:

\begin{eqnarray}
\phi_M(z) & = & \frac{1}{\mu} \int_{0}^{z} x P(x) dx\, \nonumber \\
               & = &  \frac{1}{2} \left \lbrack 1 + \mbox{\rm erf}  \left(\frac{\ln z -(m+ s^2)}{s \sqrt{2}}\right ) \right \rbrack  \, .
\label{e:lnMassInt}
\end{eqnarray}
The complements of these, $1-\phi_V(z)$ and $1-\phi_M(z)$, naturally, are the volume and mass fractions above the threshold density.  The mean density of the values of $F$ above $z$, $\eta(z)$ is:
\begin{equation}
\eta(z) =\mu \left [  \frac{1-\phi_M(z)}{1-\phi_V(z)} \right ].
\end{equation}

The inverses for equations \ref{e:lnVolInt} and \ref{e:lnMassInt}, $z = \varphi_V(\zeta), z = \varphi_M(\zeta)$,  give the upper density threshold, $\rho<z$,  corresponding to the given volume and mass fractions, $\zeta$:
\begin{eqnarray}
\varphi_V(\zeta) &=& \exp \lbrack \mbox{\bf InvNormalCDF}(\zeta, m, s) \rbrack \, ,  \\
\varphi_M(\zeta) &=&  \exp \lbrack \mbox{\bf InvNormalCDF}(\zeta, m+s^2, s) \rbrack \, , 
\end{eqnarray}
where ${\mbox{\bf InvNormalCDF}(p, \mu, \sigma)}$ is the inverse of the normal cumulative distribution function, for probability $p$, mean $\mu$, and standard deviation $\sigma$.  It is a non-linear function with no known closed form, returning the value $x$ for which the integral of the normal distribution gives a given probability $0 \leqslant p \leqslant 1$.   Note: See the function $\Phi^{-1}(p)$ in algorithm AS241, Wichura, M. J. Applied Statistics, Vol. 37, No. 3. (1988), pp. 477-484, here $x = {\mbox{\bf InvNormalCDF}(p, \mu, \sigma)} = \sigma \Phi^{-1}(p) + \mu$.  Mathematica and Excel both contain inverse normal distribution functions, although Excel breaks down for extreme values of $p$.

\begin{table}
{\small
\begin{center}
\begin{tabular}{r c c c c c}
\hline
\hline
$z$&$\phi_V(z)$&$\phi_M(z)$&$1-\phi_V(z)$&$1-\phi_M(z)$&$\eta(z)$\\
\hline
mode&0.090356&0.003713&0.909644&0.996287&1.095250\\
median&0.500000&0.090356&0.500000&0.909644&1.819288\\
mean&0.748343&0.251657&0.251657&0.748343&2.973657\\
\hline
0.001&3.54305E-06&2.78274E-09&9.99996E-01&1.00000E+00&1.00000E+00\\
0.005&5.02921E-04&1.85263E-06&9.99497E-01&9.99998E-01&1.00050E+00\\
0.01&2.79349E-03&1.98231E-05&9.97207E-01&9.99980E-01&1.00278E+00\\
0.05&5.83551E-02&1.82289E-03&9.41645E-01&9.98177E-01&1.06004E+00\\
0.1&1.46651E-01&8.43632E-03&8.53349E-01&9.91564E-01&1.16197E+00\\
0.25&3.57043E-01&4.41028E-02&6.42957E-01&9.55897E-01&1.48672E+00\\
0.5&5.60192E-01&1.17592E-01&4.39808E-01&8.82408E-01&2.00635E+00\\
0.75&6.75217E-01&1.88294E-01&3.24783E-01&8.11706E-01&2.49923E+00\\
0.9&7.22596E-01&2.27232E-01&2.77404E-01&7.72768E-01&2.78572E+00\\
1&7.48343E-01&2.51657E-01&2.51657E-01&7.48343E-01&2.97366E+00\\
1.1&7.70498E-01&2.74893E-01&2.29502E-01&7.25107E-01&3.15947E+00\\
1.5&8.34523E-01&3.57043E-01&1.65477E-01&6.42957E-01&3.88547E+00\\
2&8.82408E-01&4.39808E-01&1.17592E-01&5.60192E-01&4.76386E+00\\
5&9.69372E-01&7.03010E-01&3.06280E-02&2.96990E-01&9.69670E+00\\
10&9.91564E-01&8.53349E-01&8.43632E-03&1.46651E-01&1.73833E+01\\
20&9.98177E-01&9.41645E-01&1.82289E-03&5.83551E-02&3.20124E+01\\
50&9.99836E-01&9.87879E-01&1.64219E-04&1.21212E-02&7.38110E+01\\
100&9.99980E-01&9.97207E-01&1.98231E-05&2.79349E-03&1.40921E+02\\
200&9.99998E-01&9.99497E-01&1.85263E-06&5.02921E-04&2.71464E+02\\
500&1.00000E+00&9.99965E-01&5.43231E-08&3.54351E-05&6.52302E+02\\
1000&1.00000E+00&9.99996E-01&2.78274E-09&3.54305E-06&1.27322E+03\\
\hline
\end{tabular}
\end{center}
\caption{Integrals for $\mu = 1, \, \sigma^2 = 5$}
\label{t:s5}
}
\end{table}%

\begin{table}
{\small
\begin{center}
\begin{tabular}{r c c | r c c}
\hline
\hline
\multicolumn{3}{c}{Volume} & \multicolumn{3}{c}{Mass} \\
$\zeta$&$z =\varphi_V(\zeta)$&$\phi_M(z)$&$\zeta$&$z= \varphi_M(\zeta)$&$\phi_V(z) $\\
\hline
0.001&6.52322E-03&4.74153E-06&0.001&3.91393E-02&3.99138E-02 \\
0.005&1.29871E-02&4.53336E-05&0.005&7.79225E-02&1.07994E-01\\
0.010&1.81361E-02&1.23744E-04&0.010&1.08817E-01&1.61630E-01\\
0.050&4.51563E-02&1.42530E-03&0.050&2.70938E-01&3.79693E-01\\
0.100&7.34374E-02&4.39500E-03&0.100&4.40624E-01&5.22733E-01\\
0.250&1.65509E-01&2.20543E-02&0.250&9.93055E-01&7.46679E-01\\
0.500&4.08248E-01&9.03560E-02&0.500&2.44949E+00&9.09644E-01\\
0.750&1.00699E+00&2.53321E-01&0.750&6.04196E+00&9.77946E-01\\
0.900&2.26951E+00&4.77267E-01&0.900&1.36170E+01&9.95605E-01\\
0.950&3.69088E+00&6.20307E-01&0.950&2.21453E+01&9.98575E-01\\
0.990&9.18976E+00&8.38370E-01&0.990&5.51386E+01&9.99876E-01\\
0.999&2.55498E+01&9.60086E-01&0.999&1.53299E+02&9.99995E-01\\
\hline
\end{tabular}
\end{center}
\caption{Inverse Integrals for $\mu = 1, \, \sigma^2 = 5$}
\label{t:invs5}
}
\end{table}

Another analytical convenience of the log--normal distribution is that it can display a range of properties similar to other more commonly used functions.  For example, far from the peak of the distribution, the log--normal distribution can have powerlaw like qualities, {\em e.g.}  in $\log-\log$ space: 
\begin{equation}
\ln P(x)= -\ln x  -\ln \sqrt{2\pi}s - (\ln x - m)^2/ 2s^2 \, .
\label{e:powlerlaw}
\end{equation}
For $s^2  >>  (\ln x - m)^2$, the quadratic term is small for a large range of $x$, giving nearly straight powerlaw behaviour.

\section{A Turbulent Disk in a Potential }
\label{a:turb_disk}

In this appendix we present the detailed derivation for the distribution of the mean density of a gaseous disk in an axisymmetric potential. In so doing, we generalize the paper by \citet{strickland00a} by incorporating the effect of an isotropic turbulent velocity on the distribution of gas in the disk. We also correct an error relating to the ratio of the azimutal velocity to Keplerian velocity; this must be independent of height in the disk.

Let the density of warm gas be $\rho$ and its velocity be $v_i$. As described in \S~\ref{s:turb_disk} we express the density and velocity of the turbulent gas using a statistical approach. Dynamical variables are expressed in terms of mass-weightted mean and fluctuating components. For example the density and velocity are expressed as: 
\begin{equation}
\begin{array}{r c l r c l}
\rho &=& \bar \rho + \rho^\prime & \langle \rho^\prime \rangle &=& 0 \, ,\\
v_i &=& \tilde v_i + v_i^\prime & \langle \rho v_i^\prime \rangle &=& 0 \, .
\end{array}
\end{equation}
where the angular brackets express ensemble averages (see \citet{kuncic04a}).

If  $p$ is the thermal pressure in the disk and the gravitational potential is 
$\phi_G$. Then, the statistically averaged momentum equations for steady turbulent flow, reduce to:
\begin{equation}
\frac {\partial}{\partial x_j} (\bar \rho \tilde v_i \tilde v_j) +
\frac {\partial}{\partial x_j} \langle \rho v_i^\prime v_j^\prime \rangle 
= - \frac {\partial \bar p}{\partial x_i} 
- \bar \rho \frac {\partial \phi}{\partial x_i} 
\end{equation}
where the components $-\langle \rho v_i^\prime v_j^\prime \rangle$ represent the turbulent Reynolds stress. 

We assume isotropic turbulence in which $\langle \rho v_i^\prime v_j^\prime \rangle = \bar \rho \sigma_t^2 \delta_{ij}$ where $\sigma_t^2$ is the line of sight mean square turbulent velocity. In a disk in which the radial and vertical components of the mean velocity are zero and the azimuthal velocity is $\tilde v_\phi$, the momentum equations reduce to:
\begin{eqnarray}
\frac{\partial}{\partial z} \left( \frac {\bar \rho \sigma_t^2}{3} \right) &=& 
-\frac {\partial p}{\partial z} - \bar \rho \frac {\partial \phi_G}{\partial z}  \, , \\
\frac {\partial}{\partial r}  \left( \frac {\bar \rho \sigma_t^2}{3} \right)
- \bar \rho \frac {\tilde v_\phi^2}{r}
&=& - \frac {\partial \bar p}{\partial r} - \bar \rho \frac {\partial \phi_G}{\partial r} \, .
\end{eqnarray}
We next assume that the mean temperature of the disk is isothermal and that the turbulent velocity dispersion is constant or, slightly less restrictively, that $\sigma_g^2 = k T /\mu m + \sigma_t^2 = \hbox{constant}$. Then, the above equations become, using the dimensionless gravtiational potential, $\psi = \phi_G/\sigma_D^2$:
\begin{eqnarray}
\sigma_g^2 \frac {1}{\bar \rho} \frac {\partial \bar \rho}{\partial z} &=& 
- \sigma_D^2 \frac {\partial \psi}{\partial z} \label{e:rho_z}\\
\sigma_g^2 \frac {1}{\bar \rho} \frac {\partial \bar \rho}{\partial r} 
- \frac {\tilde v_\phi^2}{r} &=& - \sigma_D^2 \frac {\partial \psi}{\partial r}
\label{e:rho_r}
\end{eqnarray}
Differentiating equations~(\ref{e:rho_z}) and  (\ref{e:rho_r}) with respect to $r$ and $z$ respectively, and then subtracting, gives the compatibility condition:
\begin{equation}
\frac {1}{r} \frac {\partial \tilde v_\phi^2}{\partial z} = 0,
\end{equation}
so that the azimuthal velocity is a function of the cylindrical radius only.

\section{Structure functions}

In the final part of this summary we note that the Kolmogorov turbulence description was originally  analyzed in terms of eddy cascades and structure functions. The structure function of order $p$ of the variable $f(r)$ is defined by:
\begin{equation}
S_p(x, r) = \langle |f(x+r)-f(x)|^p \rangle
\end{equation}
where the angular brackets indicate an average over volume. 
For the 2nd order structure function there may be an invariant exponent $\zeta$ such that, for homogeneous, isotropic turbulence in which $S_p(x, r) = S_p(r)$
\begin{equation}
S_2(\lambda r) = \lambda^\zeta S_2(r) \, .
\end{equation}
If this condition holds, then $S_2(r)$ is a powerlaw:
\begin{equation}
S_2(r) \propto r^\zeta \, .
\end{equation}
A scaling index of $\zeta = 2/3$ was derived in the inertial subrange for ``nearly incompressible'' fluids. (\citet{Kolmogorov41a}, \citet{Kolmogorov41b}, \citet{Kolmogorov62}).  
Under some conditions (including: $1< \beta <3$ and $f(r)$ is homogeneous; see \citet{lewis04}) the $S_2(r)$ index, $\zeta$ and the $D(k)$ power-law index $\beta$ may be related, via the the Wiener-Khinchin Theorem, to give $\beta =  \zeta +1$.   With the popularity of Fourier techniques, $\beta$ is more widely used, and $\beta = 5/3$ is widely associated with the  Kolmogorov turbulence description.  We note that other indices are possible. For example $\beta = 2.0$ has been cited for a cascade of shocks, ( \citet{boldyrev04a} ).

\end{document}